\documentclass[10pt,3p,times]{elsarticle}

\usepackage{amsmath,amsfonts,amssymb}
\usepackage{amsthm}
\usepackage{bm}
\usepackage{booktabs}
\usepackage[toc]{appendix}
\usepackage[T1]{fontenc}
\usepackage{textcomp}
\usepackage{graphicx,epstopdf}
\usepackage[position=top,labelfont=normalfont,textfont=normalfont,singlelinecheck=off,justification=raggedright]{subfig}
\usepackage{floatrow}
\usepackage[dvipsnames]{xcolor}
\usepackage{setspace}
\usepackage{enumerate,etaremune}
\usepackage{tabularx,multirow}
\usepackage{framed}
\usepackage{hhline}
\usepackage[unicode,bookmarks=false]{hyperref}
\usepackage{url}
\hypersetup{
  colorlinks,
  citecolor=Blue,linkcolor=Blue,urlcolor=Blue
}
\usepackage{lineno}
\usepackage[textsize=footnotesize,linecolor=red,backgroundcolor=red!25,bordercolor=red]
  {todonotes}

\usepackage{tikz}
\usetikzlibrary{shapes.geometric}
\usetikzlibrary{shapes,arrows}
\usetikzlibrary{plotmarks}

\usetikzlibrary{calc}

\usepackage{algorithm}
\usepackage{algorithmicx,algpseudocode}
\usepackage{cases}

\newcommand{\eg}{{\it e.g.}}
\newcommand{\ie}{{\it i.e.}}

\newcommand{\etal}{{\it et al.}}
\newcommand{\tensor}[1]{\bm{#1}}
\newcommand{\stress}{\sigma}
\newcommand{\strain}{\varepsilon}
\newcommand{\tstress}{\tensor{\stress}}
\newcommand{\tstrain}{\tensor{\strain}}

\newcommand{\od}{\mathrm{d}}
\newcommand{\pd}{\partial}

\newcommand{\el}{\mathrm{e}}

\newcommand{\rn}[1]{\uppercase\expandafter{\romannumeral #1\relax}}
\newcommand{\cn}{\mathrm{N}}

\newcommand{\fric}{\mathrm{friction}}

\newcommand{\tfe}{\mathcal{G}_{\rn{1}}}
\newcommand{\sfe}{\mathcal{G}_{\rn{2}}}
\newcommand{\yield}{\mathrm{Y}}
\newcommand{\dI}{d_{\rn{1}}}
\newcommand{\dII}{d_{\rn{2}}}
\newcommand{\cdfI}{\mathcal{H}_{\rn{1}}}
\newcommand{\cdfII}{\mathcal{H}_{\rn{2}}}

\DeclareMathOperator{\grad}{\nabla}
\DeclareMathOperator{\diver}{\grad\cdot}
\DeclareMathOperator{\symgrad}{\nabla^{s}}

\DeclareMathOperator{\dyad}{\otimes}
\newsavebox{\dotbox}

\theoremstyle{remark}
\newtheorem{remark}{Remark}
\newtheorem*{remark*}{Remark}

\newcommand{\revised}[1]{{\color{black} #1}}

\newcolumntype{L}[1]{>{\raggedright\let\newline\\arraybackslash\hspace{0pt}}m{#1}}
\newcolumntype{C}[1]{>{\centering\let\newline\\arraybackslash\hspace{0pt}}m{#1}}
\newcolumntype{R}[1]{>{\raggedleft\let\newline\\arraybackslash\hspace{0pt}}m{#1}}

\allowdisplaybreaks

\biboptions{sort&compress,square,comma,numbers}
\AtBeginDocument{\hypersetup{citecolor=MidnightBlue,linkcolor=MidnightBlue,urlcolor=MidnightBlue}}

\usepackage{etoolbox}
\makeatletter
\patchcmd{\ps@pprintTitle}
{Preprint submitted to}
{~}
{}{}
\makeatother

\begin{document}

\begin{frontmatter}

\title{Double-phase-field formulation for mixed-mode fracture in rocks}

\author[HKU]{Fan Fei}
\author[HKU]{Jinhyun Choo\corref{corr}}
\ead{jchoo@hku.hk}

\cortext[corr]{Corresponding Author}

\address[HKU]{Department of Civil Engineering, The University of Hong Kong, Hong Kong}

\journal{~}

\begin{abstract}
Cracking of rocks and rock-like materials exhibits a rich variety of patterns where tensile (mode \rn{1}) and shear (mode \rn{2}) fractures are often interwoven.
\revised{These mixed-mode fractures are usually cohesive (quasi-brittle) and frictional.
Although phase-field modeling is increasingly used for rock fracture simulation, no phase-field formulation is available for cohesive and frictional mixed-mode fracture.}
To address this shortfall, here we develop a double-phase-field formulation that employs two different phase fields to describe cohesive tensile fracture and frictional shear fracture individually.
The formulation rigorously combines the two phase fields through three approaches:
(i) crack-direction-based decomposition of the strain energy into the tensile, shear, and pure compression parts,
(ii) contact-dependent calculation of the potential energy,
and (iii) energy-based determination of the dominant fracturing mode in each contact condition.
We validate the proposed model, both qualitatively and quantitatively, with experimental data on mixed-mode fracture in rocks.
\revised{The validation results demonstrate that the double-phase-field model---a combination of two quasi-brittle phase-field models---allows one to directly use material strengths measured from experiments, unlike brittle phase-field models for mixed-mode fracture in rocks.}
Another standout feature of the double-phase-field model is that it can simulate, and naturally distinguish between, tensile and shear fractures without complex algorithms.
\end{abstract}


\end{frontmatter}

\section{Introduction}
\label{sec:intro}
Rocks and rock-like materials (\eg~concrete and stiff soils) commonly fail in a quasi-brittle manner, characterized by progressive softening during the post-peak stage.
During the softening process, numerous microcracks develop, grow, and coalesce to form localized macroscopic fractures.
The region of pervasive microcracking---commonly referred to as a fracture process zone---in these materials has a non-negligible size, violating the premise of linear elastic fracture mechanics \revised{(LEFM)}.
For this reason, a number of non-linear fracture mechanics approaches have been developed and widely used for modeling the failure process in quasi-brittle materials.
Representative examples are cohesive zone models (\eg~\cite{barenblatt1962mathematical,dugdale1960yielding,needleman1987continuum,park2009unified}) and damage-type models (\eg~\cite{bazant1983crack,bazant1985microplane,pijaudier1987nonlocal,peerlings1996gradient}).

Apart from its quasi-brittleness, the cracking behavior of rocks and rock-like materials exhibits a few important characteristics.
First, these materials are fractured under compression, showing a rich variety of cracking patterns that emanate from preexisting flaws.
These rock cracking patterns often involve complex combinations of tensile (mode \rn{1}) and shear (mode \rn{2}) fractures, which have attracted a large number of experimental and numerical studies for decades (\eg~\cite{ingraffea1980finite,bobet1998fracture,bobet1998numerical,wong2009crack-a,wong2009crack-b,wong2009systematic,lee2011experimental,zhang2012cracking,yin2014coalescence,zhou2018experimental}).
Second, a sliding fracture under compressive stress entails marked friction along the crack surface.
This friction plays an important role not only in the kinematics of fracture but also in the propagation dynamics~\cite{palmer1973growth,puzrin2005growth}.
Last but not least, the shear fracture energy of rock is usually much greater than the tensile fracture energy of the same material~\cite{wong1982shear,shen1994modification}.
All these characteristics should be properly considered to accurately model cracking processes in rock.
Unfortunately, however, computational models that can efficiently simulate a combination of cohesive and frictional fractures remain scarce.

Over the past several years, phase-field modeling has gained increasing popularity for rock fracture simulation,
mainly due to its ability to capture complex crack patterns without the need for algorithmic tracking of evolving crack geometry.
The majority of phase-field simulations of rock fracture have used models that are theoretically equivalent to LEFM for brittle materials (\eg~\cite{lee2016pressure,choo2018cracking,ha2018liquid,santillan2018phase}).
However, these brittle phase-field models are not fully appropriate for rocks and rock-like materials for the reasons described above.

Meanwhile, a few studies have proposed phase-field models tailored to rocks and similar geologic materials.
The work of Zhang \etal~\cite{zhang2017modification} may be the first endeavor to modify a standard phase-field formulation for brittle fracture to distinguish between the mode \rn{1} and mode \rn{2} fracture energies of rock-like materials.
The key idea of their modification is to adopt the $\mathcal{F}$-criterion proposed by Shen and Stephansson~\cite{shen1994modification}, whereby the energy release rates of mode \rn{1} and mode \rn{2} fractures are normalized by their corresponding fracture energies.
Bryant and Sun~\cite{bryant2018mixed} later used the same idea to develop a phase-field formulation for mixed-mode fracture in anisotropic rocks.
However, these models are limited to purely brittle, pressure-insensitive fracture, neglecting softening behavior and friction effects.
Alternatively, Choo and Sun~\cite{choo2018coupled} proposed a coupled phase-field and plasticity modeling framework for pressure-sensitive geomaterials.
While this modeling framework can well simulate brittle, quasi-brittle, and ductile failures and their transitions, it does not explicitly distinguish between tensile and shear fractures.
Also importantly, the phase-field formulations underpinning all these models---originate from brittle fracture theory---inevitably suffer from a drawback that the material strength is sensitive to the length parameter for phase-field regularization.
For this reason, previous studies usually calibrated the fracture energy in conjunction with the length parameter such that their combination gives a prescribed peak stress.
However, this calibration is undesirable because the fracture energy is a material property, whereas the length parameter emanates from geometric regularization in phase-field modeling.

In recent years, a new class of phase-field models has emerged for cohesive tensile fracture.
Drawing on the gradient damage models of Lorentz and coworkers~\cite{lorentz2011conergence,lorentz2011gradient,lorentz2017nonlocal}, these phase-field models have incorporated one-dimensional softening behavior through careful design of functions for geometric regularization and material degradation.
Notable examples are the phase-field cohesive zone models advanced by Wu and coworkers~\cite{wu2017unified,wu2018length,nguyen2018modeling,feng2018phase,mandal2019length}, as well as the phase-field model for dynamic cohesive fracture by Geelen \etal~\cite{geelen2019phase}.
Apart from the explicit treatment of softening behavior, these models commonly have the feature that the material behavior is virtually insensitive to the phase-field length parameter, allowing one to use the fracture energy as a pure material parameter.
These models are thus robust and effective for simulating tensile fracture in quasi-brittle materials; however, they are not suited for shear fracture which is common in rocks.

Very recently, the first phase-field model for frictional shear fracture has been developed for geologic materials~\cite{fei2020phaseshear}.
Built on the phase-field method for frictional interfaces~\cite{fei2020phasecontact}, the new model has been derived and verified to be insensitive to the length parameter like the phase-field models for cohesive tensile fracture.
Remarkably, the new phase-field model explicitly incorporates the frictional energy into the crack propagation mechanism, in a way that is demonstrably consistent with the celebrated theory of Palmer and Rice~\cite{palmer1973growth} for frictional shear fracture.
However, the previous work restricted its attention to shear fracture, leaving its extension to mixed-mode fracture as a future research topic.

\revised{
Importantly, the phase-field formulations for cohesive tensile fracture and frictional shear fracture---derived for length-insensitive modeling of quasi-brittle behavior---cannot be combined using an existing approach to phase-field modeling of mixed-mode fracture.
The existing phase-field models for brittle mixed-mode fracture in rocks~\cite{zhang2017modification,bryant2018mixed} have commonly incorporated the difference in modes \rn{1} and \rn{2} fracture energies by replacing the standard crack driving force with an weighted average of modes \rn{1} and \rn{2} crack driving forces based on the $\mathcal{F}$-criterion.
Even though this approach has been successful for phase-field modeling of brittle fracture, it is fundamentally incompatible with that of quasi-brittle fracture.
The reason is that this approach unavoidably modifies the crack driving forces (and the degradation functions) of quasi-brittle phase-field models, which should be preserved to model the prescribed softening behavior without length sensitivity.
Therefore, for phase-field modeling of quasi-brittle mixed-mode fracture, one needs to develop a new approach that combines two length-insensitive phase-field models without altering their crack driving forces and degradation functions.
}

In this work, we propose a new phase-field formulation that employs two different phase fields to individually describe cohesive tensile fracture and frictional shear fracture for mixed-mode fracture in rocks and rock-like materials.
In the literature, multi-phase-field modeling has been used for fracture in anisotropic materials and composites (\eg~\cite{nguyen2017multi,na2018computational,bleyer2018phase,dean2020multi}), and Bleyer \etal~\cite{bleyer2018phase} have briefly suggested its application to mixed-mode fracture in brittle materials.
To our knowledge, however, no previous work has developed a multi-phase-field formulation for mixed-mode fracture in brittle materials, not to mention for mixed cohesive tensile/frictional shear fracture in quasi-brittle materials.
\revised{Critically, approaches in the existing multi-phase-models are inadequate for modeling mixed-mode fracture in rocks.
For example, the idea of overlapping multiple phase fields~\cite{nguyen2017multi,na2018computational} cannot be applied because different fracture modes should not coexist within the same material point.
Stress-based criteria used in other multi-phase-field models~\cite{bleyer2018phase,dean2020multi} cannot properly distinguish between tensile and shear fractures.}

To rigorously couple the two phase fields---one for mode \rn{1} fractures and the other for mode \rn{2}---in rocks under compression, here we devise three approaches.
First, we decompose the strain energy into the tensile, shear, and pure compression parts, based on the direction of crack at the material point.
This approach unifies the phase-field method for frictional contact~\cite{fei2020phasecontact} with the phase-field formulation for opening fracture proposed by Steinke and Kaliske~\cite{steinke2019phase}.
Second, we formulate the incremental potential energy of the material point depending on its contact condition: open, slip, or stick.
This approach extends the derivation procedure of the phase-field model for frictional shear fracture~\cite{fei2020phaseshear} to double-phase-field modeling of mixed-mode fracture.
Third, we determine the dominant fracture mode in each contact condition based on the $\mathcal{F}$-criterion for mixed-mode fracture~\cite{shen1994modification}.
Importantly, this approach is different from the way in which the $\mathcal{F}$-criterion is used in the previous single-phase-field models for mixed-mode fracture (\eg~\cite{zhang2017modification,bryant2018mixed}).
While the previous models have used the criterion to calculate an weighted average of modes \rn{1} and \rn{2} crack driving forces, here we apply it to find the dominant fracture mode and direction based on the current contact condition.
Consequently, unlike the previous single-phase-field models, the double-phase-field model clearly distinguishes between modes \rn{1} and \rn{2} fractures.

The paper is organized as follows.
In Section~\ref{sec:formulation}, we develop a double-phase-field formulation for mixed-mode fracture in quasi-brittle materials, in which one phase-field describes cohesive tensile fracture and the other phase-field describes frictional shear fracture.
This section describes the main contributions of this work.
Subsequently, Section~\ref{sec:discretization} presents discrete formulations and algorithms for numerical solution to the proposed model using the standard finite element method.
The double-phase-field model is then validated in Section~\ref{sec:validation}, both qualitatively and quantitatively, with experimental results on various mixed-mode fractures in rocks.
We conclude the work in Section~\ref{sec:closure}.


\section{Double-phase-field formulation for mixed-mode fracture}
\label{sec:formulation}
In this section, we develop a double-phase-field formulation for mixed-mode fracture in rocks and rock-like materials.
\revised{For this purpose, we apply the microforce approach in da Silva \etal~\cite{daSilva2013sharp}---adopted by Geelen \etal~\cite{geelen2019phase} and Fei and Choo~\cite{fei2020phaseshear} for deriving cohesive and frictional phase-field fracture models, respectively---to double-phase-field modeling of mixed-mode fracture.
Although the original phase-field models are formulated based on variational principles for brittle fracture (the seminal work of Francfort and Marigo~\cite{francfort1998revisiting} and its extensions), microforce theory allows one to derive phase-field models for more complex problems for which sound variational principles are unavailable, such as cohesive/frictional fracture (see Choo and Sun~\cite{choo2018coupled} for a detailed discussion).
It is noted that for the particular case of brittle fracture, the microforce and variational approaches lead to the same phase-field formulation.}
Without loss of generality, we restrict our attention to an isotropic and linear elastic material, infinitesimal deformation, rate-independent fracture, and quasi-static conditions.

\subsection{Double-phase-field approximation of tensile and shear fractures}
Consider the domain $\Omega$ with boundary $\pd \Omega$.
The boundary is decomposed into the displacement (Dirichlet) boundary $\pd_u\Omega$ and the traction (Neumann) boundary $\pd_t\Omega$, satisfying $\overline{\partial_{u}\Omega\cap\partial_{t}\Omega}=\emptyset$ and $\overline{\partial_{u}\Omega\cup\partial_{t}\Omega}=\partial\Omega$.
The domain may have two mutually exclusive sets of mode \rn{1} and mode \rn{2} fractures, which are denoted by $\Gamma_{\rn{1}}$ and $\Gamma_{\rn{2}}$, respectively.

\begin{figure}[h!]
    \centering
    \includegraphics[width = \textwidth]{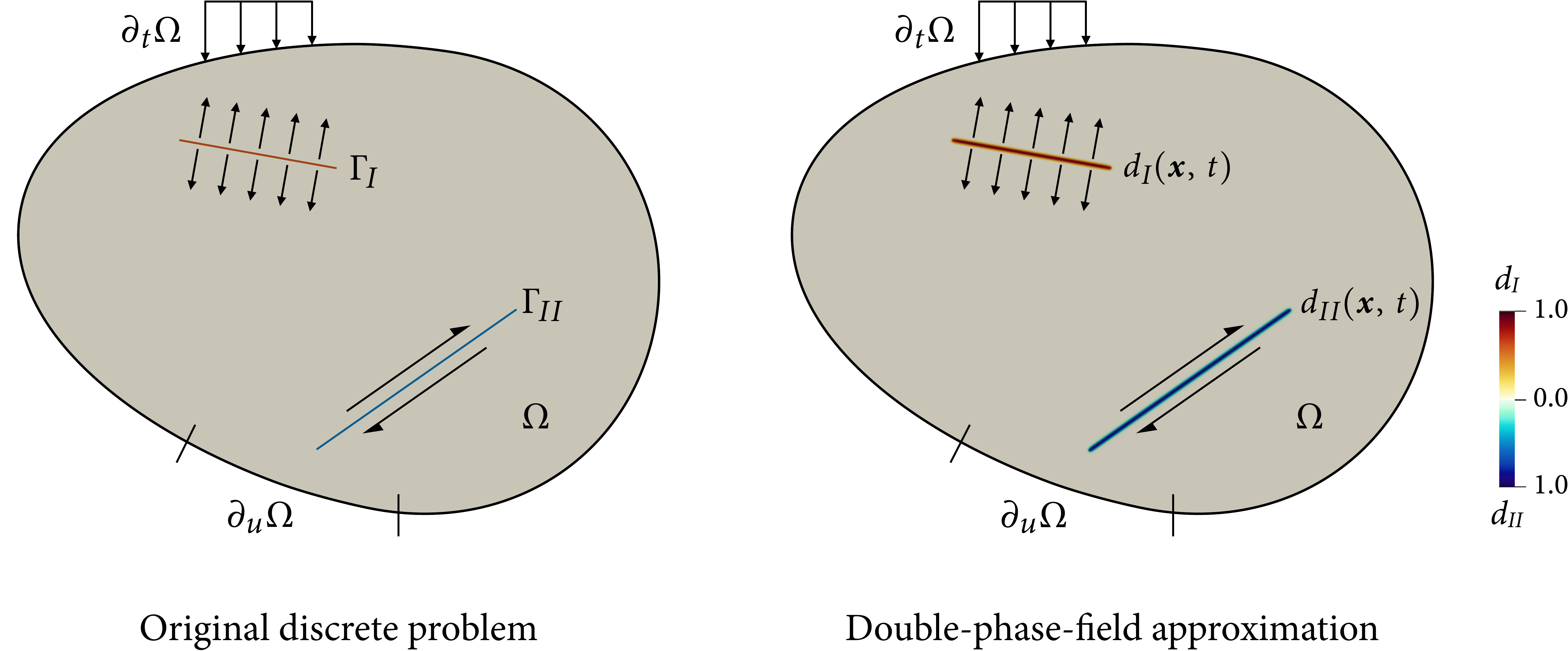}
    \caption{Double-phase-field approximation of the discontinuous geometries of mode \rn{1} (in red) and mode \rn{2} (in blue) fractures.}
    \label{fig:phase-field-approximation}
\end{figure}

To approximate the discontinuous surfaces of $\Gamma_{\rn{1}}$ and $\Gamma_{\rn{2}}$, we introduce two different phase fields: (i) $\dI$ for the mode \rn{1} fractures in $\Gamma_{\rn{1}}$, and (ii) $\dII$ for the mode \rn{2} fractures in $\Gamma_{\rn{2}}$.
Figure~\ref{fig:phase-field-approximation} illustrates this double-phase-field approximation of mixed-mode fracture.
Each of the two phase fields is defined in between 0 and 1, \ie~$\dI \in \left[0, 1\right]$ and $\dII \in \left[0, 1\right]$, such that 0 denotes an intact (undamaged) region and 1 denotes a discontinuous (fully damaged) region for the corresponding mode of fracture.

The use of two phase fields results in two crack density functions: (i) $ \Gamma_{\dI}$ for the mode \rn{1} fractures, and (ii) $\Gamma_{\dII}$ for the mode \rn{2} fractures.
For both crack density functions, we adopt the general form proposed by Wu~\cite{wu2017unified} for phase-field modeling of cohesive fracture.
Specifically,
\begin{align}
    \Gamma_{\dI}(\dI, \grad \dI) &= \dfrac{1}{\pi L} \left[ (2\dI - \dI^2) + L^2 (\grad \dI)^2 \right], \label{eq:crack-density-function-wu-mode-1}\\
    \Gamma_{\dII}(\dII, \grad \dII) &= \dfrac{1}{\pi L} \left[ (2\dII - \dII^2) + L^2 (\grad \dII)^2 \right]. \label{eq:crack-density-function-wu-mode-2}
\end{align}
Here, $L$ is the length parameter for phase-field approximation, which is assumed to be the same for both mode \rn{1} and mode \rn{2} fractures.

\subsection{Potential energy density}
To derive equations that govern the evolutions of the two phase fields, we should formulate the potential energy density of a material point.
The potential energy density, denoted by $\psi$, is decomposed into four terms~\cite{fei2020phaseshear}
\begin{align}
    \psi = \psi^\mathrm{e} + \psi^\mathrm{f} + \psi^\mathrm{d} - \psi^\mathrm{b}\,.
    \label{eq:TPE}
\end{align}
Here, $\psi^\mathrm{e}$ is the strain energy stored from elastic deformation,
$\psi^\mathrm{f}$ is the frictional energy dissipated by sliding along a crack,
$\psi^\mathrm{d}$ is the fracture energy dissipated by generation of a new crack surface,
and $\psi^\mathrm{b}$ is the external energy from body force.
Expressions for these four terms are described below.

\subsubsection*{Strain energy}
For double-phase-field modeling of fracture, we need to derive a new form of strain energy in which the two phase fields coexist.
\revised{To begin, let us consider quantities of an undamaged ($\dI=\dII=0$) material, which are often referred to as ``effective'' quantities in damage mechanics.}
The undamaged strain energy can be written as
\begin{align}
    W(\tstrain) = \dfrac{1}{2} \tstrain:\bar{\mathbb{C}}:\tstrain\,,
\end{align}
where $\tstrain$ is the infinitesimal strain tensor and $\bar{\mathbb{C}}$ is the undamaged stress-strain tangent tensor.
As the undamaged region is assumed to be isotropic and linear elastic, $\bar{\mathbb{C}}$ can be written specifically as
\begin{align}
    \bar{\mathbb{C}} = K \boldsymbol{1} \dyad \boldsymbol{1} + 2G\left(\mathbb{I} - \dfrac{1}{3}\boldsymbol{1} \dyad \boldsymbol{1} \right),
\end{align}
where $K$ and $G$ are the bulk modulus and the shear modulus, respectively,
$\tensor{1}$ is the second-order identity tensor,
and $\mathbb{I}$ is the fourth-order symmetric identity tensor.

To model mixed-mode fracture, we additively decompose the undamaged strain energy into three parts:
(i) the tensile (mode \rn{1}) part, $W^{+}_{\rn{1}}$,
(ii) the shear (mode \rn{2}) part, $W^{+}_{\rn{2}}$,
and (iii) the pure compression (non-fracturing) part, $W^{-}(\tstrain)$,
\ie
\begin{align}
    W(\tstrain) = W^{+}_{\rn{1}}(\tstrain) + W^{+}_{\rn{2}}(\tstrain) + W^{-}(\tstrain)\,.
\end{align}
This decomposition of the undamaged strain energy gives rise to the following three partial undamaged stress tensors:
\begin{align}
    \bar{\tstress}^{+}_{\rn{1}} := \frac{\pd W^+_{\rn{1}}(\tstrain)}{\pd \tstrain}\,, \quad
    \bar{\tstress}^{+}_{\rn{2}} := \frac{\pd W^+_{\rn{2}}(\tstrain)}{\pd \tstrain}\,, \quad
    \bar{\tstress}^{-} := \frac{\pd W^-(\tstrain)}{\pd \tstrain}\,.
\end{align}
By definition, the sum of the three partial undamaged stress tensors should be equal to the total undamaged stress tensor, \ie
\begin{align}
    \bar{\tstress}^{+}_{\rn{1}} + \bar{\tstress}^{+}_{\rn{2}} + \bar{\tstress}^{-}
    = \frac{\pd W(\tstrain)}{\pd \tstrain}
    \equiv \bar{\tstress} \,.
\end{align}

To calculate the specific forms of the partial undamaged stress tensors, we decompose the stress tensor with respect to the direction of the crack.
The purpose of this directional decomposition is to accommodate the phase-field model for frictional shear fracture~\cite{fei2020phaseshear}, which uses the same decomposition scheme.
The directional decomposition scheme is also compatible with opening fracture, as proposed by Steinke and Kaliske~\cite{steinke2019phase} for brittle tensile fracture.

When the directional decomposition is used, the partial undamaged stress tensors are expressed differently depending on the  contact condition of the crack: open, stick, or slip.
The contact condition can be identified following the phase-field method for frictional cracks~\cite{fei2020phasecontact}.
Let us denote by $\tensor{n}$ the unit normal vector of the crack,
by $\tensor{m}$ the unit vector in the slip direction,
and by $\tensor{s}$ the unit vector mutually orthogonal to $\tensor{n}$ and $\tensor{m}$.
The crack is open if
\begin{align}
    \strain_{nn} := \tstrain:(\tensor{n} \dyad \tensor{n}) > 0\,,
\end{align}
which corresponds to the gap condition in contact mechanics.
Equivalently, we can use the contact normal component of the undamaged stress tensor as
\begin{align}
    \bar{\stress}_{nn} := \bar{\tstress}: (\tensor{n} \dyad \tensor{n}) > 0 \,.
    \label{eq:open-condition-check-stress}
\end{align}
If the above condition is unsatisfied, the crack is closed (in contact), and it may be either in a stick or a slip condition.
To distinguish between the stick and slip conditions, we introduce a yield function of the following form:
\begin{align}
    f := |\tau| - \tau_{\yield} \leq 0\,,
    \label{eq:yield-function}
\end{align}
where
\begin{align}
    \tau := \frac{1}{2}\tstress:\bm{\alpha}\,, \;\;\mbox{with}\;\;
    \bm{\alpha} := \bm{m}\dyad\bm{n} + \bm{n}\dyad\bm{m}\,,
\end{align}
is the resolved shear stress in the crack, and $\tau_{\yield} := p_{\cn}\tan\phi$
is the yield strength, which is a function of the contact normal pressure, \revised{$p_{\cn} := -\tstress: (\tensor{n} \dyad \tensor{n})$}, and the friction angle, $\phi$.
The yield function gives $f<0$ in the stick condition and $f=0$ in the slip condition.

Depending on the contact condition, $\bar{\tstress}^{+}_{\rn{1}}$ and $\bar{\tstress}^{+}_{\rn{2}}$ are calculated as follows:
\begin{align}
    \bar{\tstress}^{+}_{\rn{1}} &=
    \left \{
        \begin{array}{ll}
        \bar{\stress}_{nn} (\tensor{n} \dyad \tensor{n}) + (\lambda/M)\bar{\stress}_{nn} [(\tensor{m} \dyad \tensor{m}) + (\tensor{s} \dyad \tensor{s})]
        &  \mbox{if}\;\; \mbox{open}\,,  \\
        \tensor{0} & \mbox{if}\;\; \mbox{stick}\,, \\
        \tensor{0} & \mbox{if}\;\; \mbox{slip}\,,
        \end{array}
    \right.
    \label{eq:mode-1-stress} \\
    \bar{\tstress}^{+}_{\rn{2}} &=
    \left \{
        \begin{array}{ll}
             \bar{\tau}\tensor{\alpha} & \mbox{if}\;\; \mbox{open}\,,\\
             \tensor{0} & \mbox{if}\;\; \mbox{stick}\,,\\
             \bar{\tau}\tensor{\alpha} & \mbox{if}\;\; \mbox{slip}\,,
        \end{array}
    \right.
    \label{eq:mode-2-stress}
\end{align}
where $M := K + (4/3) G$ is the 1D constrained modulus, $\lambda := K - (2/3) G$ is Lame's first parameter, and $\bar{\tau}:=(1/2)\bar{\tstress}:\bm{\alpha}$ is the undamaged resolved shear stress.
Also, regardless of the contact condition, $\bar{\tstress}^{-}$ is given by
\begin{align}
    \bar{\tstress}^{-} &= \bar{\tstress} - \bar{\tstress}^{+}_{\rn{1}} - \bar{\tstress}^{+}_{\rn{2}}\,.
    \label{eq:undamaged-stress}
\end{align}

Using these partial undamaged stress tensors, we write the (damaged) stress tensor, $\tstress$, as
\begin{align}
    \tstress(\tstrain,\dI,\dII)
    = g_{\rn{1}}(\dI)\bar{\tstress}^{+}_{\rn{1}}(\tstrain)
    + g_{\rn{2}}(\dII)\bar{\tstress}^{+}_{\rn{2}}(\tstrain)
    + \bar{\tstress}^{-}(\tstrain)  \,.
\end{align}
Here, $g_{\rn{1}}(\dI)\in[0,1]$ and $g_{\rn{2}}(\dII)\in[0,1]$ are the degradation functions for mode \rn{1} and mode \rn{2} fractures, respectively.
Their specific expressions will be presented later in this section.
Note that we have multiplied $g_{\rn{1}}(\dI)$ to $\bar{\tstress}^{+}_{\rn{1}}$ only,
and $g_{\rn{2}}(\dII)$ to $\bar{\tstress}^{+}_{\rn{2}}$ only.
Also importantly, the stress--strain relationship is incrementally nonlinear, because $\bar{\tstress}^{+}_{\rn{1}}$, $\bar{\tstress}^{+}_{\rn{2}}$ and $\bar{\tstress}^{-}$ are dependent on the contact condition.
Due to this incremental nonlinearity, we write the strain energy density as a rate form as
\begin{align}
  \dot{\psi}^\mathrm{e} = \left[ g_{\rn{1}}(\dI) \bar{\tstress}^{+}_{\rn{1}} + g_{\rn{2}}(\dII)\bar{\tstress}^{+}_{\rn{2}} + \bar{\tstress}^{-}  \right]:\dot{\tstrain}\,.
  \label{eq:strain-energy-density}
\end{align}

\subsubsection*{Frictional energy}
Although open cracks are frictionless, sliding cracks may involve significant friction.
This friction plays an important role in shear fracture propagation, as formally shown by Palmer and Rice~\cite{palmer1973growth}.
Therefore, the frictional energy dissipated along a sliding crack should also be incorporated into the phase-field formulation.

The frictional energy density is also an incrementally nonlinear function because frictional energy only dissipates during slip.
So we write the frictional energy density as a rate form
\begin{align}
    \dot{\psi}^\mathrm{f} = \left[1 - g_{\rn{2}}(\dII) \right] \tstress_\fric:\dot{\tstrain}\,,
    \label{eq:friction-energy-general}
\end{align}
where $\tstress_\fric$ denotes the stress tensor at the crack associated with frictional slip.
Its specific expressions, which depend on the contact condition, are given by~\cite{fei2020phasecontact}
\begin{align}
    \tstress_\fric =
    \left \{
    \begin{array}{ll}
         \tensor{0} & \mbox{if}\;\; \mbox{open}\,,  \\
         \tensor{0} & \mbox{if}\;\; \mbox{stick}\,, \\
         \tau_{r} \tensor{\alpha} & \mbox{if}\;\; \mbox{slip}\,.
    \end{array}
    \right.
    \label{eq:friction-stress}
\end{align}
Here, $\tau_{r}$ is the residual shear strength of the fracture, which equals $\tau_{\mathrm{Y}}$ during slip.
Inserting Eq. \eqref{eq:friction-stress} into Eq.~\eqref{eq:friction-energy-general}, we obtain the rate form of frictional energy density as
\begin{align}
    \dot{\psi}^\mathrm{f} =
    \left \{
    \begin{array}{ll}
         0 & \mbox{if}\;\; \mbox{open}\,,  \\
         0 & \mbox{if}\;\; \mbox{stick}\,, \\
         \left[1 - g_{\rn{2}}(\dII) \right] \tau_{r} \dot{\gamma} & \mbox{if}\;\; \mbox{slip}\,,
    \end{array}
    \right.
    \label{eq:friction-energy-density}
\end{align}
where $\gamma := \tstrain:\tensor{\alpha}$ denotes the shear strain at the crack.

\subsubsection*{Fracture energy}
Since we consider two different modes of fracture, the fracture energy dissipation is additionally decomposed into two terms as
\begin{align}
    \psi^\mathrm{d} = \psi^\mathrm{d}_{\rn{1}} + \psi^\mathrm{d}_{\rn{2}}\,,
\end{align}
where $\psi^\mathrm{d}_{\rn{1}}$ and $\psi^\mathrm{d}_{\rn{2}}$ correspond to energy dissipation densities associated with modes \rn{1} and \rn{2} fractures, respectively.
Let $\tfe$ and $\sfe$ denote the critical fracture energies for mode \rn{1} and \rn{2} fractures.
Then the two terms can be expressed as
\begin{align}
    \psi^\mathrm{d}_{\rn{1}} &= \tfe\Gamma_{\dI} = \dfrac{\tfe}{\pi L} \left[(2\dI - \dI^2) + L^2 (\grad \dI)^2 \right],  \label{eq:dissipation-density-mode-1} \\
    \psi^\mathrm{d}_{\rn{2}} &= \sfe\Gamma_{\dII} =  \dfrac{\sfe}{\pi L} \left[(2\dII - \dII^2) + L^2 (\grad \dII)^2 \right]. \label{eq:dissipation-density-mode-2}
\end{align}

\subsubsection*{External energy}
The external energy, which is due to gravitational force, can be written as
\begin{align}
    \psi^\mathrm{b} = \rho \tensor{g} \cdot \tensor{u}\,,
\end{align}
where $\rho$ is the mass density, $\tensor{g}$ is the gravitational acceleration vector,
and $\bm{u}$ is the displacement vector.

\subsection{Governing equations}
According to the microforce argument~\cite{daSilva2013sharp}, the governing equations of the problem are obtained as follows:
\begin{align}
    &\diver\, \left(\dfrac{\pd \psi(\tstrain,\dI, \grad \dI, \dII, \grad \dII)}{\pd \tstrain} \right)
    - \dfrac{\pd \psi(\tstrain,\dI, \grad \dI, \dII, \grad \dII) }{\pd \tensor{u}}
    = \tensor{0}
    && \mbox{(momentum balance)}\,,
    \label{eq:momentum-balance-eq}\\
    &\diver\, \left(\dfrac{\pd \psi(\tstrain,\dI, \grad \dI, \dII, \grad \dII)}{\pd \grad \dI} \right)
    - \dfrac{\pd \psi(\tstrain,\dI, \grad \dI, \dII, \grad \dII)}{\pd \dI}
    = -\pi_{r,\rn{1}}
    && \mbox{(mode \rn{1} microforce balance)}\,,
    \label{eq:microforce-balance-eq-mode-1} \\
    &\diver\, \left(\dfrac{\pd \psi(\tstrain,\dI, \grad \dI, \dII, \grad \dII)}{\pd \grad \dII} \right)
    - \dfrac{\pd \psi(\tstrain,\dI, \grad \dI, \dII, \grad \dII)}{\pd \dII}
    = -\pi_{r,\rn{2}}
    && \mbox{(mode \rn{2} microforce balance)}\,.
    \label{eq:microforce-balance-eq-mode-2}
\end{align}
Here, $\pi_{r,{\rn{1}}}$ and $\pi_{r,{\rn{2}}}$ are reactive microforces introduced to ensure the irreversibility of modes \rn{1} and \rn{2} fracture processes, respectively.
Their specific expressions will be presented later in this section, after other terms become derived.

Substituting the previously derived expressions for the potential energy into Eqs.~\eqref{eq:momentum-balance-eq}, \eqref{eq:microforce-balance-eq-mode-1}, and \eqref{eq:microforce-balance-eq-mode-2},
we get more specific forms of the governing equations as
\begin{align}
    \diver\, \tensor{\tstress} + \rho\tensor{g} &= \tensor{0}\,,
    \label{eq:momentum-balance-eq-specific} \\
    - g'_{\rn{1}}(\dI)\cdfI
    - \dfrac{\tfe}{\pi L} \left( 2L^2 \diver \grad \dI - 2 + 2\dI \right) &=
    -\pi_{r,\rn{1}}\,,
    \label{eq:microforce-balance-eq-mode-1-specific} \\
    - g'_{\rn{2}}(\dII)\cdfII
    - \dfrac{\sfe}{\pi L} \left( 2L^2 \diver \grad \dII - 2 + 2\dII \right)
    &= -\pi_{r,\rn{2}}\,.
    \label{eq:microforce-balance-eq-mode-2-specific}
\end{align}
Here, $\cdfI$ and $\cdfII$ are the (undamaged) crack driving forces for modes \rn{1} and \rn{2} fractures, respectively,
which are related to the derivatives of the potential energy with respect to the two phase fields.
Because the potential energy has been formulated differently according to the contact condition,
the crack driving forces must be dependent on the current contact condition.

\subsection{Modes \rn{1} and \rn{2} crack driving forces under different contact conditions}
A unique challenge for double-phase-field modeling of mixed mode fracture is to prevent overlapping of modes \rn{1} and \rn{2} within the same material point.
This requires careful determination of the modes \rn{1} and \rn{2} crack driving forces, $\cdfI$ and $\cdfII$ at every material point.
To this end, here we adapt the $\mathcal{F}$-criterion proposed by Shen and Stephansson~\cite{shen1994modification} to the double-phase-field modeling of mixed-mode fracture.
Defining $\theta$ as the angle between the crack normal direction and the major principal direction in the slip plane, we rephrase the idea of the $\mathcal{F}$-criterion as:
\begin{align}
    \theta = \arg \max_{\theta} \left[\mathcal{F}(\theta)\right] \rvert_{\tstrain}\,,
    \;\;\mbox{with}\;\;
    \mathcal{F}(\theta) := \dfrac{\cdfI(\tstrain,\theta)}{\tfe} + \dfrac{\cdfII(\tstrain,\theta)}{\sfe} \, .
    \label{eq:F-criterion-phase-field}
\end{align}
In other words, the mixed-mode fracture propagates such that the value of $\mathcal{F}$ is maximized.
It is noted that the strain tensor, $\tstrain$, in the argument may be replaced by the undamaged stress tensor, $\bar{\tstress}$, as $\tstrain=\bar{\mathbb{C}}^{-1}:\bar{\tstress}$.

Based on the foregoing derivations of the potential energy and the $\mathcal{F}$-criterion, we derive specific forms of the modes \rn{1} and \rn{2} crack driving forces in the following four cases:
(i) the intact (undamaged) condition, (ii) the open condition, (iii) the stick condition, and (iv) the slip condition.

\subsubsection*{Intact condition}
Let us first consider an intact material point in which neither mode \rn{1} nor \rn{2} fracture has yet developed.
To prevent fracturing in the elastic region, we set $\cdfI$ and $\cdfII$ as their threshold values defined as the crack driving forces at the peak tensile and shear strengths, respectively.
Let $\mathcal{H}_{{\rn{1}}, t}$ denote the threshold for $\cdfI$.
By definition, $\mathcal{H}_{{\rn{1}}, t}$ corresponds to the undamaged tensile strain energy, $W^{+}_{\rn{1}}$, when $\bar{\stress}_{nn}$ equals the tensile strength.
Thus we get
\begin{align}
    \mathcal{H}_{{\rn{1}}, t} := W^{+}_{\rn{1}}|_{\bar{\stress}_{nn}=\stress_{p}}
    = \dfrac{1}{2}\bar{\tstress}^{+}_{\rn{1}}|_{\bar{\stress}_{nn}=\stress_{p}} : \tstrain = \dfrac{1}{2M} \stress_{p}^{2}\,,
    \label{eq:H-threhsold-mode-1}
\end{align}
where $\stress_{p}$ denotes the tensile strength.
The derivation of $\mathcal{H}_{{\rn{2}}, t}$ is more complex and long due to the existence of frictional energy dissipation in shear fracture.
Referring to Fei and Choo~\cite{fei2020phaseshear} for a detailed derivation of $\mathcal{H}_{{\rn{2}}, t}$, we adopt
\begin{align}
    \mathcal{H}_{{\rn{2}},t} := \dfrac{1}{2G}(\tau_{p} - \tau_{r})^2\,,
    \label{eq:H-threhsold-mode-2}
\end{align}
where $\tau_{p}$ is the peak shear strength.
Because the threshold values are assigned as the crack driving forces of an intact material point,
\begin{align}
  \left.\begin{array}{ll}
    \cdfI = \dfrac{1}{2M} \stress_{p}^{2} \\ [0.5em]
    \cdfII = \dfrac{1}{2G}(\tau_{p} - \tau_{r})^2
  \end{array}\right\}
  \;\text{if}\;\; \text{intact}\,.
\end{align}
In this work, we treat $\stress_{p}$ as a constant material property, but consider $\tau_{p}$ a function of the contact normal pressure.
Specifically, we set $\tau_{p} = c_{0} + p_{\cn}\tan \phi$, where $c_{0}$ and $\phi$ denote the cohesion and the friction angle, respectively.
For simplicity, we assume that the peak and residual friction angles are the same and calculate the residual shear strength as $\tau_{r}=p_{\cn}\tan\phi$.
This assumption can be easily relaxed as in Fei and Choo~\cite{fei2020phaseshear}.

\subsubsection*{Open condition}
Next, we consider the case in which a crack develops under an open contact condition.
To determine the dominant fracturing mode in this case, we need to evaluate the value of $\mathcal{F}$ given in Eq.~\eqref{eq:F-criterion-phase-field}, and hence $\cdfI$ and $\cdfII$ therein.
To this end, we only have to consider the strain energy density, $\psi^{\el}$, because an open crack is frictionless ($\psi^{\mathrm{f}}=0$) and other energy terms ($\psi^{\mathrm{d}}$ and $\psi^{\mathrm{b}}$) are unrelated to the crack driving forces.
To compute $\psi^\mathrm{e}$ during post-peak fracturing, we integrate its rate form in Eq. \eqref{eq:strain-energy-density} from the peak stresses, as
\begin{align}
	\psi^\mathrm{e} = \dfrac{1}{2}\bar{\tstress}^{-}:\tstrain + g_{\rn{1}}(\dI)  \left[ W^{+}_{\rn{1}} \bigr \rvert_{t_{p,{\rn{1}}}} + \dfrac{1}{2}\left(\bar{\tstress}^{+}_{\rn{1}}:\tstrain \right) \bigr \rvert^{t}_{t_{p,{\rn{1}}}} \right] + g_{\rn{2}}(\dII) \left[ W^{+}_{\rn{2}} \bigr \rvert_{t_{p,{\rn{2}}}} + \dfrac{1}{2} \left(\bar{\tstress}^{+}_{\rn{2}}:\tstrain \right) \bigr \rvert^{t}_{t_{p,{\rn{2}}}}\right] \, .
  \label{eq:strain-energy-open-loading-integral}
\end{align}
Here, $t_{p,{\rn{1}}}$ and $t_{p,{\rn{2}}}$ denote the time instances when $\stress_{p}$ and $\tau_{p}$ are reached, respectively, and
\begin{align}
	W^{+}_{\rn{1}}\rvert_{t_{p,{\rn{1}}}} = \dfrac{1}{2M} \stress^2_{p} \, , \quad
	W^{+}_{\rn{2}}\rvert_{t_{p,{\rn{2}}}} = \dfrac{1}{2G} \tau^2_{p} \, ,
\end{align}
are the strain energies relevant to modes \rn{1} and \rn{2} fracturing, respectively, at the corresponding peak stresses.
Plugging the above expressions and Eqs.~\eqref{eq:mode-1-stress},~\eqref{eq:mode-2-stress}, and \eqref{eq:undamaged-stress} into Eq. \eqref{eq:strain-energy-open-loading-integral}, we obtain
\begin{align}
	\psi^\mathrm{e}
  = \dfrac{1}{2} \bar{\tstress}:\tstrain
  - [1 - g_{\rn{1}}(\dI)] \dfrac{1}{2M}\bar{\stress}^2_{nn}
  - [1 -  g_{\rn{2}}(\dII)] \dfrac{1}{2G}\bar{\tau}^2 \, .
  \label{eq:strain-energy-open-loading}
\end{align}
By definition, $\cdfI$ and $\cdfII$ must be the terms multiplied to $[1 - g_{\rn{1}}(\dI)]$ and $[1 - g_{\rn{2}}(\dII)]$, respectively.
Therefore,
\begin{align}
	\mathcal{H}_{\rn{1}} &= \dfrac{1}{2M}\bar{\stress}^2_{nn} \,,
  \label{eq:H1-open-loading}\\
	\mathcal{H}_{\rn{2}} &= \dfrac{1}{2G}\bar{\tau}^2 \, .
  \label{eq:H2-open-loading}
\end{align}
Substituting Eqs.~\eqref{eq:H1-open-loading} and~\eqref{eq:H2-open-loading} into the $\mathcal{F}$-criterion~\eqref{eq:F-criterion-phase-field}, we get
\begin{align}
  \mathcal{F}(\tstrain, \theta) = \dfrac{\bar{\stress}^{2}_{nn}}{2M \tfe} + \dfrac{\bar{\tau}^2}{2G\sfe} \, .  \label{eq:F-open}
\end{align}
In terms of the principal strains and $\theta$, the above equation can be re-written as
\begin{align}
    \mathcal{F}(\tstrain, \theta) = \dfrac{\left[\lambda\left(\strain_1 \sin^2 \theta  + \strain_3 \cos^2 \theta \right) + M\left(\strain_1\cos^2 \theta + \strain_3 \sin^2 \theta \right)  \right]^2}{2M\tfe} + \dfrac{2G\left(\strain_1 - \strain_3 \right)^2\cos^2\theta \sin^2 \theta}{\sfe} \, ,  \label{eq:F-open-principal}
\end{align}
where $\strain_1$ and $\strain_3$ denote the major and minor principal strains.
Then, to find $\theta$ that maximizes $\mathcal{F}$, we take the partial derivative of $\mathcal{F}$ with respect to $\theta$ as
\begin{align}
    \dfrac{\pd \mathcal{F}(\tstrain, \theta)}{\pd \theta}
    = \dfrac{2G\left(\lambda +  G \right)}{M \tfe} \left(\strain^2_3 - \strain^2_1 \right)\sin 2\theta + G(\strain_1 - \strain_3)^2 \sin 4\theta \left(\dfrac{1}{\sfe} - \dfrac{G}{M\tfe} \right) \, .
\end{align}
This derivative becomes zero when $\theta = 0$.
This means that under an open condition, the crack should develop such that the crack normal direction is the same as the major principal direction.
When the crack direction is determined in this way, $\mathcal{H}_{{\rn{2}}}=0$ because no shear stress exists on the principal plane, \ie~$\bar{\tau}=0$ when $\theta=0$.
Therefore,
\begin{align}
  \left.\begin{array}{ll}
    \cdfI = \dfrac{1}{2M} \bar{\stress}_{nn}^{2} \\ [0.5em]
    \cdfII = 0
  \end{array}\right\}
  \;\text{if}\;\; \text{open}\,.
\end{align}
In other words, a mode \rn{2} crack does not grow (\ie~$\dot{\dII}=0$) under open conditions.
Note that in this case, $\bar{\stress}_{nn}$ should be equal to the major principal undamaged stress, $\bar{\stress}_{1}$, because $\theta=0$.

\subsubsection*{Stick condition}
We now shift our focus to a material point that has a closed crack under a stick condition.
In this case, $\bar{\tstress}^{+}_{\rn{1}}=\bar{\tstress}^{+}_{\rn{2}}=0$,
thus $\pd\psi^\mathrm{e}/\pd \dI=\pd\psi^\mathrm{e}/\pd \dII=0$.
The frictional energy is zero as well (\ie~$\psi^{\mathrm{f}}$).
Therefore,
\begin{align}
  \left.\begin{array}{ll}
    \cdfI = 0 \\ [0.5em]
    \cdfII = 0
  \end{array}\right\}
  \;\text{if}\;\; \text{stick}\,.
\end{align}
So neither mode \rn{1} nor mode \rn{2} crack grows (\ie~$\dot{\dI}=0$ and $\dot{\dII}=0$) when there is no relative motion between the two crack surfaces.
This result is also physically intuitive because a material with a perfectly sticky crack behaves like an undamaged material.

\subsubsection*{Slip condition}
Lastly, we consider the case when the material point has a closed crack undergoing slip.
In this case, it can be easily shown that $\cdfI=0$, because $\pd\psi^\mathrm{e}/\pd \dI=0$ and $\pd\psi^\mathrm{f}/\pd \dI=0$ when the crack is sliding.
Therefore, maximizing $\mathcal{F}$ in Eq.~\eqref{eq:F-criterion-phase-field} is equivalent to maximizing $\cdfII$ in slip conditions.
This means that $\cdfII$ can be derived in the exact same way as in the phase-field model for shear fracture~\cite{fei2020phaseshear}.
Below we briefly recap the derivation of $\cdfII$, referring to Fei and Choo~\cite{fei2020phaseshear} for details.
We first evaluate the strain energy and frictional energy densities by integrating their rate forms given in Eqs. \eqref{eq:strain-energy-density} and \eqref{eq:friction-energy-density}, and get
\begin{align}
    \psi^\mathrm{e}
    &= \psi^\mathrm{e}\rvert_{t_{p}} + \int^{t}_{t_{p}} \bar{\tstress}^{-} : \dot{\tstrain} \, \od t
    + \int^{t}_{t_{p}} g_{\rn{2}}(\dII)  \bar{\tstress}^{+}_{\rn{2}} \, \od t \, ,
    \label{eq:strain-energy-density-slip} \\
    \psi^\mathrm{f}
    &= \int^{t}_{t_{p}}[1 - g_{\rn{2}}(\dII)] \tau_{r} \dot{\gamma} \, \od t \, ,
    \label{eq:friction-energy-density-slip}
\end{align}
Taking the partial derivatives of $\psi^\mathrm{e}$ and $\psi^\mathrm{f}$ with respect to $\dII$, we obtain $\cdfII$ as
\begin{align}
    \cdfII = \mathcal{H}_{\rn{2},t} + \mathcal{H}_\mathrm{slip}\,,
    \;\; \mbox{with} \;\;
    \mathcal{H}_\mathrm{slip} := \int_{\gamma_{p}}^\gamma (\bar{\tau} - \tau_{r}) \: \od \gamma \, .
    \label{eq:slip-H-mode-2}
\end{align}
Here, $\mathcal{H}_\mathrm{slip}$ denotes the crack driving force accumulated during the post-peak slip process, and $\gamma_{p}$ is the shear strain in the slip direction when $\tau = \tau_{p}$.
We note that $\mathcal{H}_\mathrm{slip}$ is expressed as an integral form because $\tau_{r}$ is a function of the contact normal pressure.
Now, we determine the crack propagation direction by maximizing $\mathcal{F}$, equivalently, $\cdfII$, in this case.
As derived in Fei and Choo~\cite{fei2020phaseshear}, it eventually boils down to find $\theta$ such that
\begin{align}
    \theta = \arg \max_\theta [\bar{\tau}(\theta) - \tau_{r}(\theta)] \, .
\end{align}
When $\tau_{r}=p_{\cn}\tan\phi$, we get
\begin{align}
    \theta = 45^\circ - \dfrac{\phi}{2}
    \label{eq:theta-slip} \,,
\end{align}
see Fei and Choo~\cite{fei2020phaseshear} for details.
Note that this value of $\theta$ is necessary to calculate $\bar{\tau}$ and $\gamma$ in $\mathcal{H}_\mathrm{slip}$.
To summarize,
\begin{align}
  \left.\begin{array}{ll}
    \cdfI = 0 \\ [0.5em]
    \cdfII =  \dfrac{1}{2G}(\tau_{p} - \tau_{r})^2 + \displaystyle\int_{\gamma_{p}}^\gamma (\bar{\tau} - \tau_{r}) \: \od \gamma
  \end{array}\right\}
  \;\text{if}\;\; \text{slip}\,.
\end{align}
As opposed to the previous case of open fracture, a mode \rn{1} crack does not grow (\ie~$\dot{\dI}=0$) in the slip case.
This result also agrees well with our physical intuition.
\smallskip

\revised{
\begin{remark}
  The present double-phase-field model applies the $\mathcal{F}$-criterion~\cite{shen1994modification} in a largely different way from how previous single-phase-field models (\eg~\cite{zhang2017modification,bryant2018mixed}) have used it for mixed-mode fracture.
  In Zhang \etal~\cite{zhang2017modification}, the $\mathcal{F}$-criterion is used to calculate an equivalent crack driving force as an weighted average of modes \rn{1} and \rn{2} crack driving forces.
  With the same equivalent crack driving force, Bryant and Sun~\cite{bryant2018mixed} have further used the $\mathcal{F}$-criterion to determine the fracture direction by solving an optimization problem at the material point level.
  However, instead of calculating an equivalent crack driving force, here we apply the $\mathcal{F}$-criterion to determine the dominant fracture mode (phase field) and its evolution direction.
  The upshot is that the double-phase-field model not only distinguishes between modes \rn{1} and \rn{2} fractures naturally but also calculates the fracturing direction based on the $\mathcal{F}$-criterion without solving an optimization problem.
\end{remark}
}

\subsection{Crack irreversibility}
Having derived the modes \rn{1} and \rn{2} crack driving forces under all contact conditions,
we can now specify expressions for the modes \rn{1} and \rn{2} reactive microforces, $\pi_{r,{\rn{1}}}$ and $\pi_{r,{\rn{2}}}$, which prevent spurious crack healing.
Assuming that both modes \rn{1} and \rn{2} cracks do not heal at all,
here we set the reactive microforces following the method proposed by Miehe~\etal~\cite{miehe2010phase} whereby the crack driving force is replaced by the maximum crack driving force in loading history.
\revised{Despite being simple, this history-based method has been shown to provide results fairly similar to those obtained by a more sophisticated and robust algorithm---see, \eg~its comparison with an augmented Lagrangian method in Geelen \etal~\cite{geelen2019phase}.
The simplicity and effectiveness of the history-based method is more appreciable for double-phase-field modeling in which two phase fields are subjected to irreversibility constraints.}

Applying the history-based method, we define the reactive forces for modes \rn{1} and modes \rn{2} fracture as
\begin{align}
    \pi_{r,\rn{1}} &=
    \left\{\begin{array}{ll}
      0 & \text{if} \;\; \revised{\dot{\dI}}>0\,,\\ [0.5em]
      -g'_{\rn{1}}(\dI)\displaystyle\max_{t \in [0,t]} \cdfI(t) + g'_{\rn{1}}(\dI) \cdfI & \mbox{if} \;\; \revised{\dot{\dI}}=0\,,
    \end{array}\right.
    \label{eq:reactive-force-mode1}\\
    \pi_{r,\rn{2}} &=
    \left\{\begin{array}{ll}
      0 & \mbox{if} \;\; \dot{\dII}>0\,,\\ [0.5em]
      -g'_{\rn{2}}(\dII)\displaystyle\max_{t \in [0,t]} \cdfII(t)\ & \mbox{if} \;\; \dot{\dII}=0\,.
    \end{array}\right.
    \label{eq:reactive-force-mode2}
\end{align}
where $t$ denotes the current time instance.
Note that the last term in Eq.~\eqref{eq:reactive-force-mode1} is added because $\cdfI$ can be positive under an open condition.
Eq.~\eqref{eq:reactive-force-mode2} \revised{is} the same as that in the phase-field model for frictional shear fracture~\cite{fei2020phaseshear}.

\subsection{Degradation functions for modes \rn{1} and \rn{2} fractures}
To complete the formulation, we introduce specific forms to the degradation functions for modes \rn{1} and \rn{2} fractures, $g_{\rn{1}}(\dI)$ and $g_{\rn{2}}(\dII)$, respectively.
Particularly, we adopt $g_{\rn{1}}(\dI)$ from the phase-field model for cohesive tensile fracture~\cite{wu2017unified}, given by
\begin{align}
    g_{\rn{1}}(\dI) = \dfrac{\left(1 - \dI \right)^n}{\left(1 - \dI \right)^n + m_{\rn{1}}\dI\left(1 - p \dI \right)} \, ,
     \;\;\mbox{with}\;\;
     m_{\rn{1}} := \dfrac{\tfe}{\pi L}\dfrac{1}{\mathcal{H}_{\rn{1},t}} \, ,
\end{align}
and $g_{\rn{2}}(\dII)$ from the phase-field model for frictional shear fracture~\cite{fei2020phaseshear}, given by
\begin{align}
    g_{\rn{2}}(\dII) = \dfrac{\left(1 - \dII \right)^n}{\left(1 - \dII \right)^n + m_{\rn{2}}d\left(1 - p \dII \right)} \, ,
    \;\;\mbox{with}\;\;
    m_{\rn{2}} := \dfrac{\sfe}{\pi L}\dfrac{1}{\mathcal{H}_{\rn{2},t}} \, ,
\end{align}
where $n$ and $p$ are parameters controlling post-peak softening responses.
In this work, we use a standard choice of $n = 2$ and $p = -0.5$.

\section{Discretization and algorithms}
\label{sec:discretization}
In this section, we describe how to numerically solve the proposed double-phase-field formulation using \revised{a nonlinear finite element method.}

\subsection{Unified expressions for crack driving forces considering crack irreversibility}
To simplify the succeeding formulations, let us first unify the expressions for the crack driving forces and the reactive microforces under different contact conditions.
We begin this by merging the intact condition into either the open or the stick condition.
When $\bar{\stress}_{nn}> 0$, the intact condition can be combined with the open condition, because $\cdfI=\mathcal{H}_{\rn{1},t}$ initially.
Likewise, when $\bar{\stress}_{nn} \leq 0$, the intact condition can be integrated with the stick condition, as $\cdfII=\mathcal{H}_{\rn{2},t}$ initially.
These intact and stick conditions can be distinguished from the slip condition based on the value of $f$, by setting $\tau_\mathrm{Y}$ in $f$ as follows: $\tau_\mathrm{Y}=\tau_{p}$ for an intact material, and $\tau_\mathrm{Y} = \tau_{r}$ for a damaged material.
This way allows us to identify all possible conditions based on the values of $\bar{\stress}_{nn}$ and $f$.

Then, we define the combined crack driving and reactive forces for mode \rn{1} and \rn{2} fractures, $\mathcal{H}_{\rn{1}}^{+}$ and $\mathcal{H}_{\rn{2}}^{+}$, respectively, as
\begin{align}
    \mathcal{H}_{\rn{1}}^{+} &=
    \left \{
    \begin{array}{ll}
         \max \left \{\mathcal{H}_{\rn{1},t}, \, \dfrac{1}{2M} \left[ \displaystyle \max_{t \in [0,t]} \bar{\stress}_{nn}(t) \right]^2 \right\}  & \mbox{if} \;\; \bar{\stress}_{nn} > 0 \, ,  \\ [0.5 em]
         \displaystyle\max_{t \in [0,t]} \cdfI(t) & \mbox{if} \; \; \bar{\stress}_{nn} \leq 0 \; \;  \mbox{and} \; \; f < 0 \, , \\ [0.5 em]
         \displaystyle\max_{t \in [0,t]} \cdfI(t)  & \mbox{if} \;\; \bar{\stress}_{nn} \leq 0 \;\;  \mbox{and} \;\; f = 0 \,,
    \end{array}
    \right.
    \label{eq:H-mode-1}
    \\
    \mathcal{H}_{\rn{2}}^{+} &=
    \left \{
    \begin{array}{ll}
         \displaystyle\max_{t \in [0,t]} \cdfII(t) & \mbox{if} \;\; \bar{\stress}_{nn} > 0 \, , \\ [0.5 em]
         \displaystyle\max_{t \in [0,t]} \cdfII(t) & \mbox{if} \; \; \bar{\stress}_{nn} \leq 0 \; \;  \mbox{and} \; \; f < 0 \, ,  \\ [0.5 em]
         \mathcal{H}_{{\rn{2}},t} + \mathcal{H}_\mathrm{slip} & \mbox{if} \;\; \bar{\stress}_{nn} \leq 0 \;\;  \mbox{and} \;\; f = 0 \,.
    \end{array}
    \right.
    \label{eq:H-mode-2}
\end{align}
Note that we update $\mathcal{H}_{\rn{1}}^{+}$ only when $\bar{\stress}_{nn} > 0$, and $\mathcal{H}_{\rn{2}}^{+}$ only when $\bar{\stress}_{nn} \leq 0$ and $f=0$.

\subsection{Problem statement}
Let us denote by $\hat{\tensor{u}}$ and $\hat{\tensor{t}}$ the prescribed displacement and traction boundary conditions, respectively, and by $\tensor{u}_0$, $d_{\rn{1}0}$ and $d_{\rn{2}0}$ the initial displacement field and the initial mode \rn{1} and \rn{2} phase fields, respectively.
The time domain is denoted by $\mathbb{T}:=(0,t_{\mathrm{max}}]$.
The strong form of the problem can then be stated as follows: find $\tensor{u}$, $\dI$ and $\dII$ that satisfy
\begin{align}
    \diver \tstress + \rho \tensor{g} &= \tensor{0} \quad \mbox{in} \quad \Omega \times \mathbb{T} \, , \\
    -g'_{\rn{1}} (\dI) \mathcal{H}_{\rn{1}}^{+} + \dfrac{\tfe}{\pi L} \left(2L^2 \diver \grad \dI - 2 + 2\dI \right) &= 0 \quad \mbox{in} \quad \Omega \times \mathbb{T} \, , \\
    -g'_{\rn{2}} (\dII) \mathcal{H}_{\rn{2}}^{+} + \dfrac{\sfe}{\pi L} \left(2L^2 \diver \grad \dII - 2 + 2\dII \right) &= 0 \quad \mbox{in} \quad \Omega \times \mathbb{T} \, ,
\end{align}
subject to boundary conditions
\begin{align}
    \tensor{u} = \hat{\tensor{u}} \quad &\mbox{on} \quad \pd_{u} \Omega \times \mathbb{T} \,, \\
    \tstress \cdot \tensor{v} = \hat{\tensor{t}} \quad &\mbox{on} \quad \pd_{t} \Omega \times \mathbb{T} \,, \\
    \grad \dI \cdot \tensor{v} = 0 \quad &\mbox{on} \quad \pd \Omega \times \mathbb{T} \,, \\
     \grad \dII \cdot \tensor{v} = 0 \quad &\mbox{on} \quad \pd \Omega \times \mathbb{T} \,,
\end{align}
with $\tensor{v}$ denoting the outward unit normal vector at the boundary,
and initial conditions
\begin{align}
    \tensor{u} \rvert_{t = 0} = \tensor{u}_0 \quad &\mbox{in} \quad \overline{\Omega}\,, \\
    \dI\rvert_{t = 0} = d_{{\rn{1}}0} \quad &\mbox{in} \quad \overline{\Omega}\,, \\
    \dII \rvert_{t = 0} = d_{{\rn{2}}0} \quad &\mbox{in} \quad \overline{\Omega}\,,
\end{align}
where $\overline{\Omega} := \overline{\Omega \cup \pd \Omega}$.

\subsection{Finite element discretization}
To begin finite element discretization, we define the trial function spaces for $\tensor{u}$, $\dI$ and $\dII$ as
\begin{align}
    \mathcal{S}_u &:= \left \{ \tensor{u} \; \rvert \; \tensor{u} \in H^1, \, \tensor{u} = \hat{\tensor{u}} \; \mbox{on} \; \pd_u \Omega \right\} , \\
    \mathcal{S}_{\dI} &:= \left \{\dI \; \rvert \; \dI \in H^1 \right\} , \\
    \mathcal{S}_{\dII} &:= \left \{\dII \; \rvert \; \dII \in H^1 \right \} ,
\end{align}
where $H^{1}$ denotes a Sobolev space of order one.
Accordingly, the weighting function spaces are defined as
\begin{align}
    \mathcal{V}_{u} &:= \left\{\tensor{\eta} \; \rvert \; \tensor{\eta} \in H^1, \, \tensor{\eta} = \tensor{0} \;\mbox{on} \; \pd_u \Omega  \right\} , \\
    \mathcal{V}_{\dI} &:= \left\{\phi_{\rn{1}} \; \rvert \; \phi_{\rn{1}} \in H^1 \right\} , \\
    \mathcal{V}_{\dII} &:= \left\{\phi_{\rn{2}} \; \rvert \; \phi_{\rn{2}} \in H^1 \right\} .
\end{align}
Applying the standard weighted residual procedure, we obtain the following variational equations:
\begin{align}
    R_{u} &:=  - \int_\Omega \symgrad \tensor{\eta} : \tstress \: \od V + \int_\Omega \rho \tensor{\eta} \cdot \tensor{g} \: \od V + \int_{\pd_t \Omega} \tensor{\eta} \cdot \hat{\tensor{t}} \: \od A = 0 \, ,
    \label{eq:variational-momentum} \\
    R_{\dI} &:=  \int_\Omega \phi_{\rn{1}} g'_{\rn{1}} (\dI) \mathcal{H}_{\rn{1}}^{+} \: \od V + \int_\Omega \dfrac{\tfe}{\pi L} \left(2L^2 \grad \phi_{\rn{1}} \cdot \grad \dI + 2\phi_{\rn{1}} - 2\phi_{\rn{1}}\dI\right) \od V = 0 \, ,
    \label{eq:variational-microforce-mode-1} \\
    R_{\dII} &:=  \int_\Omega \phi_{\rn{2}} g'_{\rn{2}} (\dII) \mathcal{H}_{\rn{2}}^{+} \: \od V + \int_\Omega \dfrac{\sfe}{\pi L} \left(2L^2 \grad \phi_{\rn{2}} \cdot \grad \dII + 2\phi_{\rn{2}} - 2\phi_{\rn{2}}\dII\right) \od V = 0 \, .
    \label{eq:variational-microforce-mode-2}
\end{align}
Here, we have defined the variational equations as residuals to solve them using Newton's method.
The rest of the finite element procedure is straightforward; so we omit it for brevity.
The standard linear elements are used for all the field variables.

\subsection{Solution strategy}
To solve the discrete versions of variational equations \eqref{eq:variational-momentum}, \eqref{eq:variational-microforce-mode-1}, and \eqref{eq:variational-microforce-mode-2}, we use a staggered scheme which has commonly been used since proposed by Miehe \etal~\cite{miehe2010phase}.
Specifically, we first solve Eq.~\eqref{eq:variational-momentum} for $\tensor{u}$ fixing $\dI$ and $\dII$,
then update the crack driving forces $\mathcal{H}_{\rn{1}}^{+}$ and $\mathcal{H}_{\rn{2}}^{+}$,
and finally solve Eqs.~\eqref{eq:variational-microforce-mode-1} and \eqref{eq:variational-microforce-mode-2} for $\dI$ and $\dII$ fixing $\mathcal{H}_{\rn{1}}^{+}$ and $\mathcal{H}_{\rn{2}}^{+}$.
Provided that the load step size is small enough, this staggered scheme significantly improves the robustness of numerical solution without much compromise in the solution accuracy.
\revised{Also, a single staggered iteration may be sufficient for practical purposes, as will be demonstrated later through a numerical example.
It is noted that other multi-phase-field models~\cite{nguyen2017multi,na2018computational,bleyer2018phase,dean2020multi} have also used staggered solution schemes.}

\revised{Because the formulation is incrementally nonlinear, we use Newton's method to solve each stage in a staggered iteration.
To solve for $\bm{u}$ in the first stage, we linearize Eq.~\eqref{eq:variational-momentum} as ($\delta$ denoting the linearization operator)
\begin{align}
  \delta R_{u} = \int_\Omega \symgrad \bm{\eta} : \mathbb{C} : \symgrad \delta\bm{u} \: \od V\,,
\end{align}
and to solve for $\dI$ and $\dII$ in the second stage, we linearize Eqs.~\eqref{eq:variational-microforce-mode-1} and~\eqref{eq:variational-microforce-mode-2} as
\begin{align}
  \delta R_{\dI} &= \int_\Omega \phi_{\rn{1}} g''_{\rn{1}} (\dI) \mathcal{H}_{\rn{1}}^{+}\, \delta \dI \: \od V
  + \int_\Omega \dfrac{\tfe}{\pi L} \left(2L^2 \grad \phi_{\rn{1}} \cdot \grad \delta \dI - 2\phi_{\rn{1}}\delta\dI \right) \od V\,, \\
  \delta R_{\dII} &= \int_\Omega \phi_{\rn{2}} g''_{\rn{2}} (\dII) \mathcal{H}_{\rn{2}}^{+}\, \delta \dII \: \od V
  + \int_\Omega \dfrac{\sfe}{\pi L} \left(2L^2 \grad \phi_{\rn{2}} \cdot \grad \delta \dII - 2\phi_{\rn{2}}\delta\dII \right) \od V\,.
\end{align}
It is noted that $\mathcal{H}_{\rn{1}}^{+}$ and $\mathcal{H}_{\rn{2}}^{+}$ are not linearized because they are fixed during the phase-field solution stage.}

Algorithm \ref{algo:material-update} presents a procedure to update internal variables at a material/quadrature point \revised{during a Newton iteration.}
Here, known quantities at the previous load step are denoted with subscript $(\cdot)_{n-1}$,
whereas \revised{unknown quantities requiring updates} are written without an additional subscript for brevity.
The procedure essentially extends the predictor--corrector algorithm of the phase-field model for shear fracture~\cite{fei2020phaseshear} to accommodate the open contact condition.
Importantly, one can see that the present model treats all the contact conditions without any algorithm for imposing contact constraints.
This feature is the main advantage of the double-phase-field model from the numerical viewpoint.

Several aspects of the algorithm may deserve elaboration.
First, the crack driving forces, $\cdfI^{+}$ and $\cdfII^{+}$, of an initially intact material point ($d_{{\rn{1}}0}=d_{{\rn{2}}0}=0$) should be initialized by their threshold values, $\mathcal{H}_{\rn{1},t}$ and $\mathcal{H}_{\rn{2},t}$, respectively, to prevent fracturing in the elastic region.
Second, because the potential fracture direction is unknown \textit{a priori}, we first evaluate $\theta$ using the undamaged major principal stress, $\bar{\stress}_{1}$ (Line 2), considering that $\bar{\stress}_{1}=\bar{\stress}_{nn}$ under an open condition.
Third, the stress tensor under a slip condition is obtained by enforcing $f=0$ (Line 20), similar to the return mapping algorithm in plasticity.
Fourth, in Line 20, the residual strength, $\tau_{r}$, is evaluated explicitly from the previous time step, as in the frictional shear fracture model~\cite{fei2020phaseshear}.
This semi-implicit update greatly simplifies the stress--strain tangent, $\mathbb{C}$, without much compromise in accuracy.
Lastly, unlike the original algorithm for shear fracture~\cite{fei2020phaseshear},
$g_{\rn{2}}(\dII)$ is not updated when $\dII = 0$ and $f<0$.
This is because the friction angle for the peak and residual strengths are assumed to be the same in this work.
If the peak and residual friction angles are considered different, $g_{\rn{2}}(\dII)$ needs to be updated as explained in Fei and Choo~\cite{fei2020phaseshear}.
This modification is straightforward.
\begin{algorithm}[h!]
    \setstretch{1.25}
    \caption{Material point update procedure for the double-phase-field model for mixed-mode fracture}
      \begin{algorithmic}[1]
      \Require $\tstrain$, $\dI$ and $\dII$.
      \Ensure $\tstress$, $\mathbb{C}$, $\mathcal{H}^{+}_{\rn{1}}$ and $\mathcal{H}^{+}_{\rn{2}}$.
      \State Calculate $\bar{\tstress} = \bar{\mathbb{C}}:\tstrain$ and $\bar{\stress}_{1}$.
      \State Set $\theta = 0^\circ$ if $\bar{\stress}_{1} > 0$; otherwise, set $\theta = 45^\circ - \phi/2$.
      \State Calculate $\tensor{n}$, $\tensor{m}$, and $\tensor{s}$ from $\theta$.
      \State Calculate $\tensor{\alpha}$ from $\tensor{n}$ and $\tensor{m}$.
      \State Calculate $\bar{\stress}_{nn} = \bar{\tstress}:\left(\tensor{n} \dyad \tensor{n} \right)$.
      \If {$\bar{\stress}_{nn} > 0$}
          \State Open condition.
          \State Update $\tstress = \bar{\tstress} - \left[ 1 - g_{\rn{1}}(\dI) \right]\{\bar{\stress}_{nn} (\tensor{n} \dyad \tensor{n}) + (\lambda/M)\bar{\stress}_{nn} [(\tensor{m} \dyad \tensor{m}) + (\tensor{s} \dyad \tensor{s})]\}$.
          \State Update $\mathbb{C} = \bar{\mathbb{C}} - \left[1 - g_{\rn{1}}(\dI) \right] \left\{ \left(\tensor{n} \dyad \tensor{n} \right)+ \left(\lambda /M \right)[ (\tensor{m} \dyad \tensor{m} ) + (\tensor{s} \dyad \tensor{s} )]\right\} \dyad \left\{ M\left(\tensor{n} \dyad \tensor{n} \right)  + \lambda [(\tensor{m} \dyad \tensor{m} ) + (\tensor{s} \dyad \tensor{s} )]  \right\}$.
          \State Update $\mathcal{H}^{+}_{\rn{1}} = \max \left[\bar{\stress}^2_{nn}/(2M),\, \left(\mathcal{H}^{+}_{\rn{1}} \right)_{n-1} \right]$.
          \State Set $\mathcal{H}^{+}_{\rn{2}} = \left( \mathcal{H}^{+}_{\rn{2}} \right)_{n-1}$.
      \Else
          \State Calculate $\bar{\tau} = (1/2)\bar{\tstress}:\tensor{\alpha}$ and $p_{\cn} = - \bar{\stress}_{nn}$.
          \State Set $\tau_\mathrm{Y} = c_{0} + p_{\cn} \tan \phi$ if $\dII = 0$; otherwise, set $\tau_\mathrm{Y} = p_{\cn} \tan \phi$.
          \State Evaluate $f = \lvert \bar{\tau} \rvert - \tau_\mathrm{Y}$.
          \If {$f < 0$}
              \State Stick condition.
              \State Update $\tstress = \bar{\tstress}$.
              \State Update $\mathbb{C} = \bar{\mathbb{C}}$.
              \State Set $\mathcal{H}^{+}_{\rn{2}} = \left( \mathcal{H}^{+}_{\rn{2}} \right)_{n-1}$.
          \Else
              \State Slip condition.
              \State Update $\tstress = \bar{\tstress} - [1 - g_{\rn{2}}(\dII)][\bar{\tau} - (\tau_{r})_{n-1}]\tensor{\alpha}$, where $(\tau_{r})_{n-1} := (p_{\cn})_{n-1} \tan\phi$.
              \State Update $\mathbb{C} = \bar{\mathbb{C}} - [1 - g_{\rn{2}}(\dII)]G(\tensor{\alpha}\dyad\tensor{\alpha})$.
              \State Update $\mathcal{H}^{+}_{\rn{2}} = \left(\mathcal{H}^{+}_{\rn{2}}\right)_{n-1} + (\bar{\tau} - \tau_r)\Delta \gamma$, where $\tau_r = p_{\cn} \tan\phi$ and $\Delta\gamma := (\tstrain - \tstrain_{n-1}):\tensor{\alpha}$.
          \EndIf
          \State Set $\mathcal{H}^{+}_{\rn{1}} = \left( \mathcal{H}^{+}_{\rn{1}} \right)_{n-1}$.
      \EndIf
      \end{algorithmic}
  \label{algo:material-update}
\end{algorithm}

\section{Validation}
\label{sec:validation}
In this section, we validate the proposed double-phase-field model with experimental data on mixed-mode fracture in rocks.
Before simulating mixed-mode fracture, we have verified that the double-phase-field model degenerates into a cohesive tensile model and a frictional shear model under pure mode \rn{1} and mode \rn{2} problems, respectively.
These verification results are omitted for brevity.
Also, we do not repeat discussions pertaining to the numerical aspects of the original phase-field models combined in this work (\eg~mesh and length sensitivity); we refer to Wu~\cite{wu2017unified} and Fei and Choo~\cite{fei2020phaseshear} for discussions on such topics.
By doing so, we fully focus on new aspects that arise from the double-phase-field formulation for mixed-mode fracture.

To validate the model, we simulate the uniaxial compression tests of Wong~\cite{wong2008}, Bobet and Einstein~\cite{bobet1998fracture} and Wong and Einstein~\cite{wong2009crack-a} on gypsum specimens with preexisting flaw(s), whereby various mixed-mode cracking patterns are characterized under different flaw configurations.
Emulating the experimental setup, we consider 76.2 mm wide and 152.4 mm tall rectangular specimens with a single or double flaws.
The flaw configuration of each specimen will be described later.

Table~\ref{tab:double-flaws-parameters} presents the material parameters used in the simulation.
Among these parameters, the elasticity parameters ($K$ and $G$) and the tensile strength ($\stress_p$) are directly adopted from their values measured from the gypsum specimens in Bobet and Einstein~\cite{bobet1998fracture}.
The cohesion strength ($c_0$) and the friction angle ($\phi$) are unavailable from the original experiment, so they are assigned referring to other experiments on molded gypsum specimens~\cite{wei2020physical}.
The tensile and shear fracture energies ($\tfe$ and $\sfe$) are calibrated to match the coalescence stresses measured in Bobet and Einstein~\cite{bobet1998fracture}.
The calibrated values and the mode mixity ratio ($\sfe/\tfe$) lie within the ranges of their typical values for rocks~\cite{shen1994modification}.
\begin{table}[h!]
    \centering
    \begin{tabular}{lllrl}
    \toprule
    Parameter & Symbol & Units & Value & Reference \\
    \midrule
    Bulk modulus & $K$ & GPa & 2.84 & Measured in~\cite{bobet1998fracture} \\
    Shear modulus & $G$ & GPa & 2.59 &  Measured in~\cite{bobet1998fracture} \\
    Tensile strength & $\stress_p$ & MPa & 3.2 &  Measured in~\cite{bobet1998fracture} \\
    Cohesion strength & $c_{0}$ & MPa & 10.7 &  Measured in~\cite{wei2020physical} \\
    Friction angle & $\phi$ & deg & 28 & Measured in~\cite{wei2020physical} \\
    Mode \rn{1} fracture energy & $\tfe$ & J/m$^{2}$ & 16 & Calibrated from data in~\cite{bobet1998fracture} \\
    Mode \rn{2} fracture energy & $\sfe$ & J/m$^{2}$ & 205 & Calibrated from data in~\cite{bobet1998fracture} \\
    \bottomrule
    \end{tabular}
    \caption{Cracking from preexisting flaws: material parameters.}
    \label{tab:double-flaws-parameters}
\end{table}

For finite element simulation, we set the phase-field length parameter as $L = 0.2$ mm and refine elements near the preexisting flaw(s) such that their size $h$ satisfies $L/h \geq 5$.
\revised{Each specimen is then discretized by around 300,000 quadrilateral elements. (The specific number depends on the flaw configuration.)}
The simulation begins by applying a constant displacement rate of $2\times 10^{-3}$ mm on the top boundary.
The bottom boundary is supported by rollers except for the left corner which is fixed by a pin for stability.
The lateral boundaries are traction free.
Gravity is ignored.
The finite element solutions are obtained using a parallel finite element code for geomechanics~\cite{choo2015stabilized,choo2018large,choo2019stabilized}, which is built on the \verb|deal.II| finite element library~\cite{dealII,dealII91}, \verb|p4est| mesh handling library~\cite{p4est}, and the \verb|Trilinos| project~\cite{trilinos}.

\revised{
\subsection{Cracking from a single flaw}
To begin, we simulate the cracking process in a single-flawed gypsum specimen, following the experimental setup in Wong~\cite{wong2008}.
Figure~\ref{fig:single-flaw-setup} illustrates the geometry and boundary conditions of the problem.
The flaw is 12.7 mm long, 1.27 mm wide, and inclined $45^\circ$ from the horizontal.
\begin{figure}[h!]
    \centering
    \includegraphics[width=0.3\textwidth]{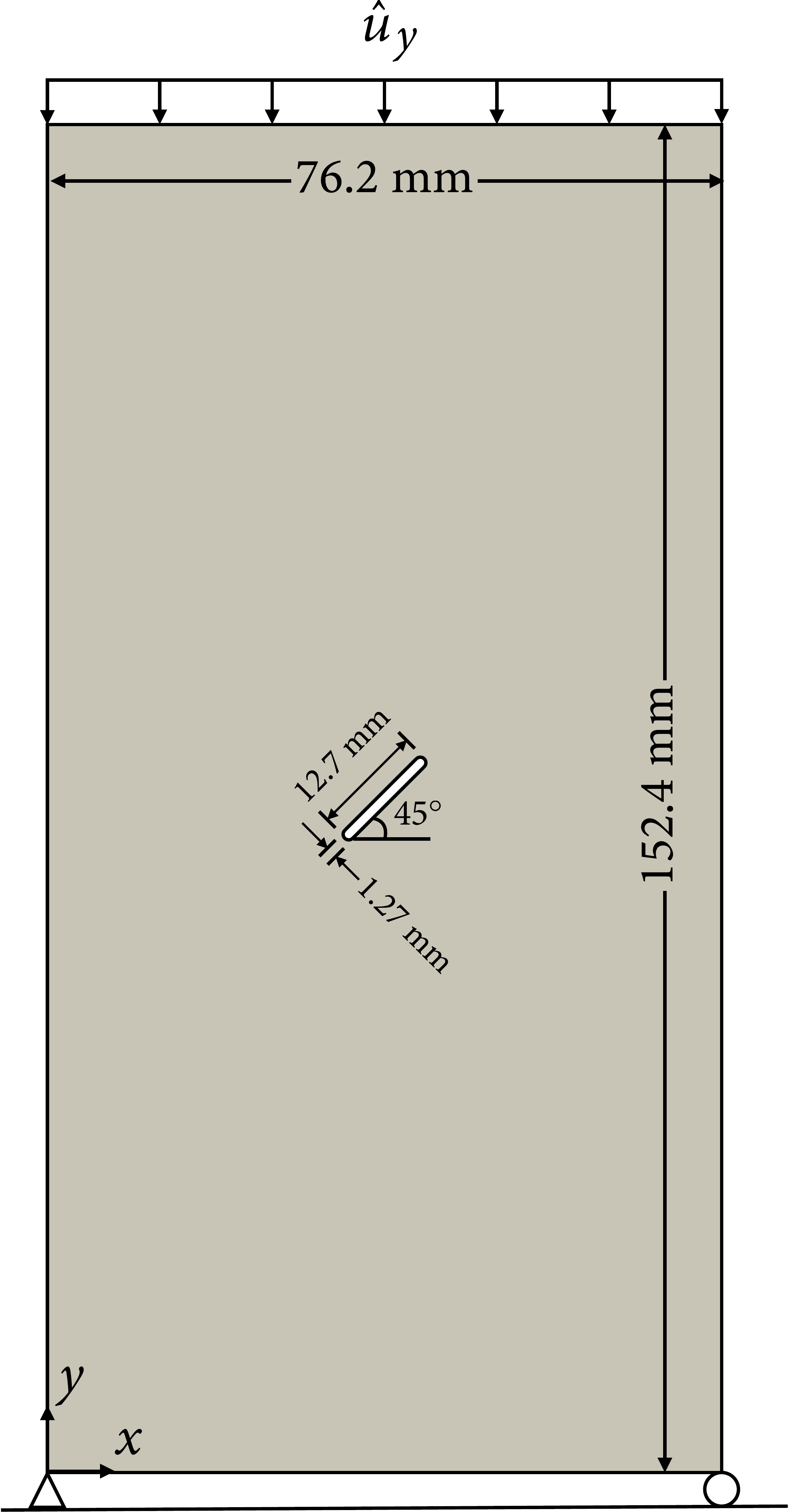}
    \caption{Cracking from a single flaw: problem geometry and boundary conditions.}
    \label{fig:single-flaw-setup}
\end{figure}

Figure~\ref{fig:single-flaw-results} presents simulation results in comparison with the cracking pattern of a specimen studied in Wong~\cite{wong2008}.
It can be seen that the double-phase-field model well reproduces the real cracking process.
When $\hat{u}_{y}=-0.40$ mm, tensile wing cracks start to grow from the flaw tips, and later at $\hat{u}_{y}=-0.60$ mm, tensile and shear damages appear.
These tensile and shear damages soon develop into full cracks at $\hat{u}_{y}=-0.66$ mm.
The final cracking pattern in our numerical simulation is nearly the same as the experimental observation.
\begin{figure}[h!]
    \centering
    \includegraphics[width=1.0\textwidth]{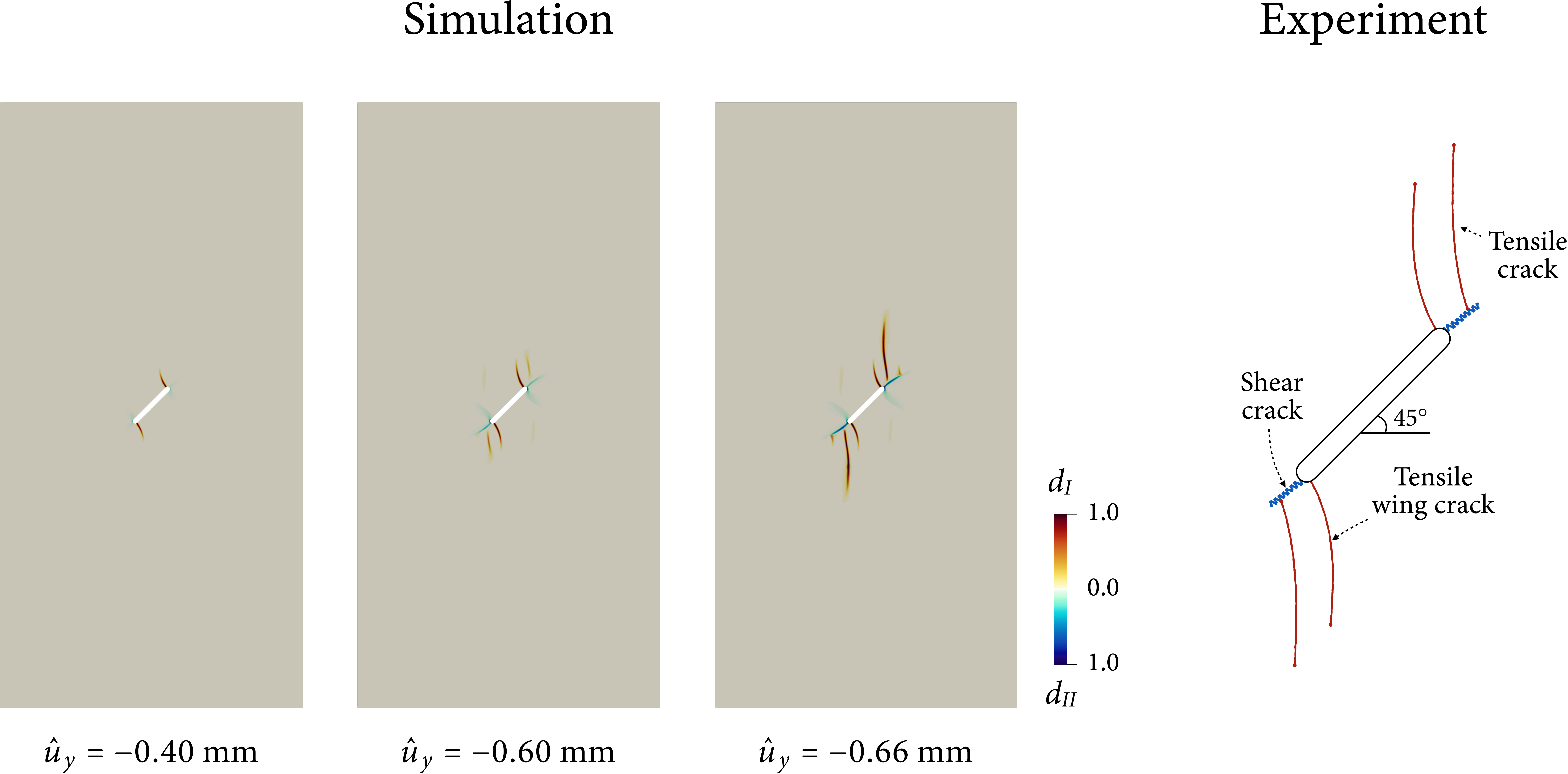}
    \caption{Cracking from a single flaw: simulation and experimental results. The experimental result is redrawn from Wong~\cite{wong2008}.}
    \label{fig:single-flaw-results}
\end{figure}

For quantitative validation, Fig.~\ref{fig:single-flaw-force-disp} compares the stress--strain curve from numerical simulation with the experimental data of Wong~\cite{wong2008} provided by the author.
The stress and strain in the specimen are defined in a nominal manner following the experimental data.
The simulation result matches remarkably well with the experimental data, even though none of the material parameters has been calibrated from this particular experiment.
Thus, the double-phase-field model has been fully validated, both qualitatively and quantitatively, with the experiment.
\begin{figure}[h!]
    \centering
    \includegraphics[width=0.55\textwidth]{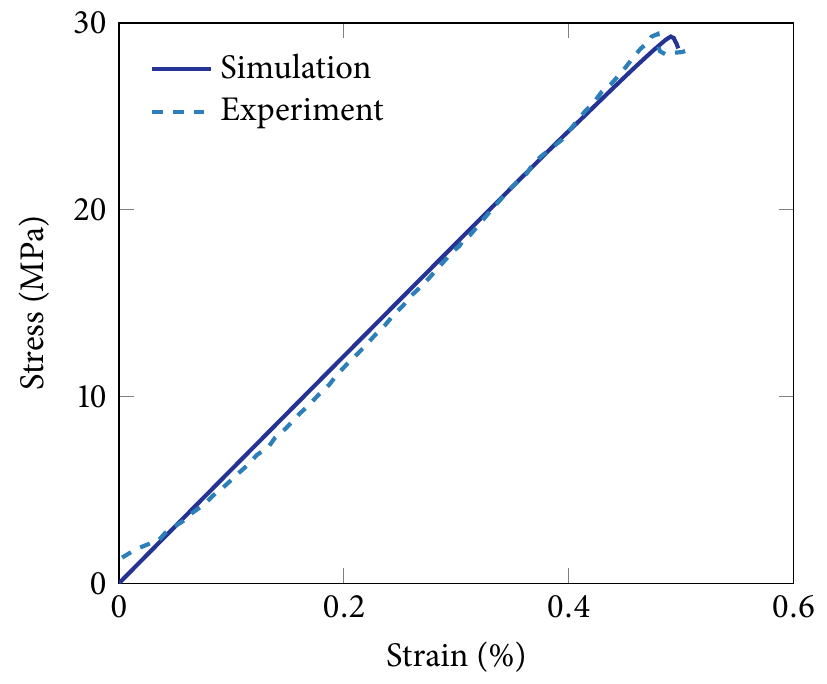}
    \caption{Cracking from a single flaw: comparison of the stress--strain curve from numerical simulation with the experimental data of Wong~\cite{wong2008} provided by the author.}
    \label{fig:single-flaw-force-disp}
\end{figure}

To strengthen the validity of our numerical results, we repeat the same simulation with different numbers of staggered iterations and compare results in Fig.~\ref{fig:single-flaw-staggered}.
One can see that the simulation results are virtually insensitive to the number of staggered iterations, in both qualitative and quantitative senses.
It can thus be concluded that as long as the load step size is chosen to be reasonably small, a single iteration is sufficiently accurate.
\begin{figure}[h!]
    \centering
    \includegraphics[width=1.0\textwidth]{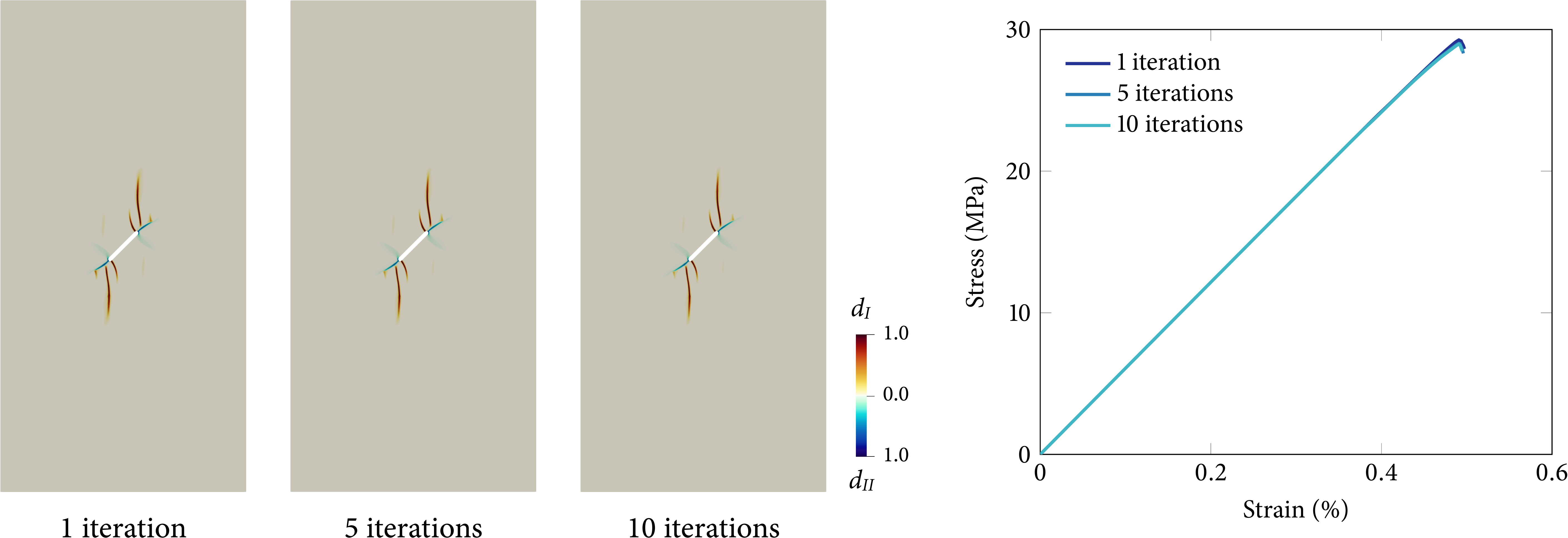}
    \caption{Cracking from a single flaw: comparison of simulation results obtained with different numbers of staggered iterations. The phase fields are drawn at $\hat{u}_{y}=-0.66$ mm.}
    \label{fig:single-flaw-staggered}
\end{figure}

Before proceeding to other validation examples, we also demonstrate why the double-phase-field model is an essential extension of previous single-phase-field models for quasi-brittle materials~\cite{wu2017unified,fei2020phaseshear} to simulate mixed-mode fracture in rocks.
Figure~\ref{fig:single-flaw-comparison} compares simulation results of the same problem obtained by the present double-phase-field model, the single-phase-field model for cohesive tensile fracture~\cite{wu2017unified}, and the the single-phase-field model for frictional shear fracture~\cite{fei2020phaseshear}.
Clearly, the single-phase-field models cannot reproduce the experimentally-observed cracking pattern presented in Fig.~\ref{fig:single-flaw-results}, even in a qualitative manner.
Thus the present model is a critical achievement for phase-field modeling of mixed-mode fracture in quasi-brittle rocks and other similar materials.
\begin{figure}[h!]
    \centering
    \includegraphics[width=0.8\textwidth]{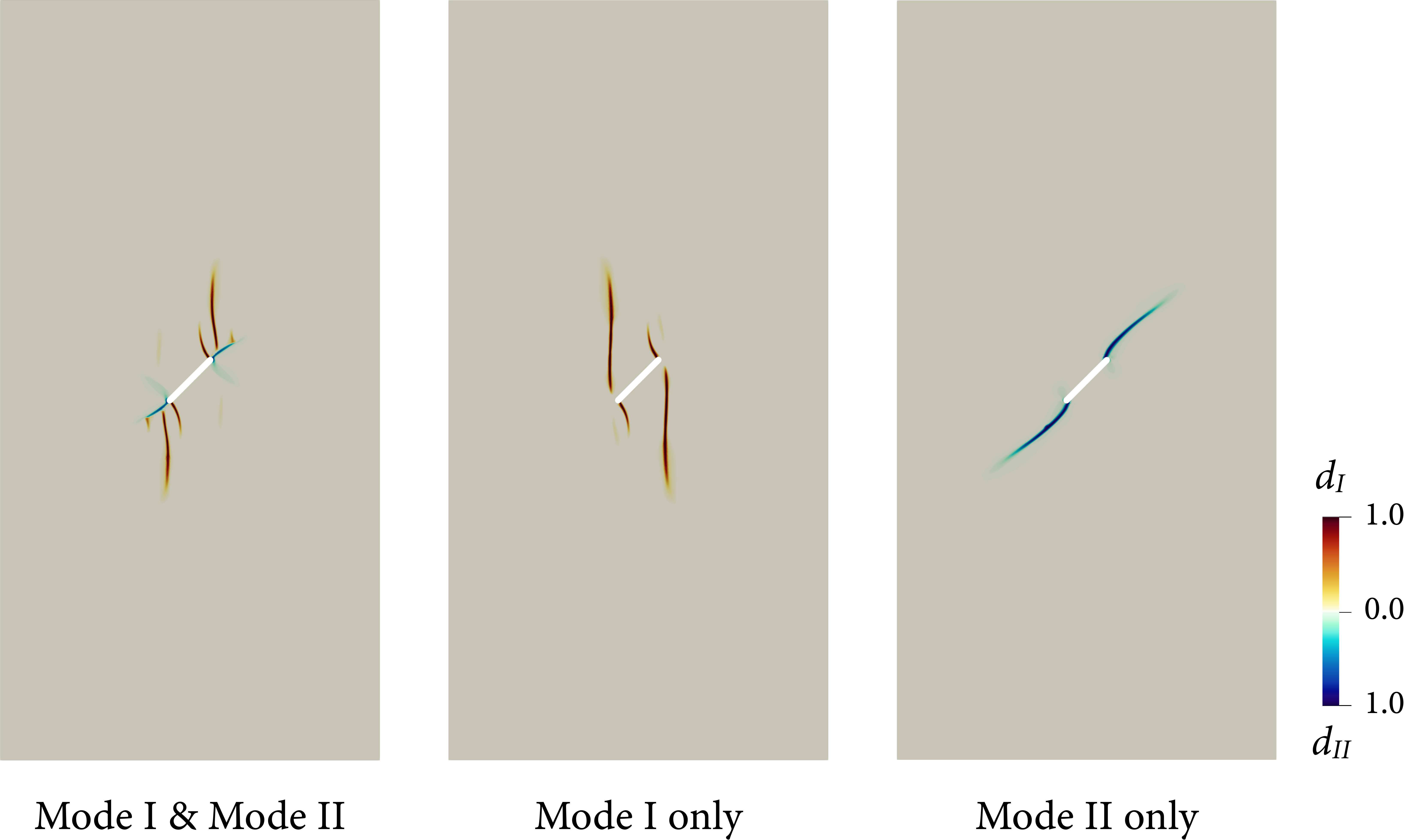}
    \caption{Cracking from a single flaw: comparison of simulation results obtained by the double-phase-field model (Mode \rn{1} \& Mode \rn{2}), the single-phase-field model for cohesive tensile fracture~\cite{wu2017unified} (Mode \rn{1} only), and the single-phase-field model for frictional shear fracture~\cite{fei2020phaseshear} (Mode \rn{2} only).}
    \label{fig:single-flaw-comparison}
\end{figure}
}

\subsection{Cracking from double flaws}
\begin{figure}[h!]
    \centering
    \includegraphics[width=0.5\textwidth]{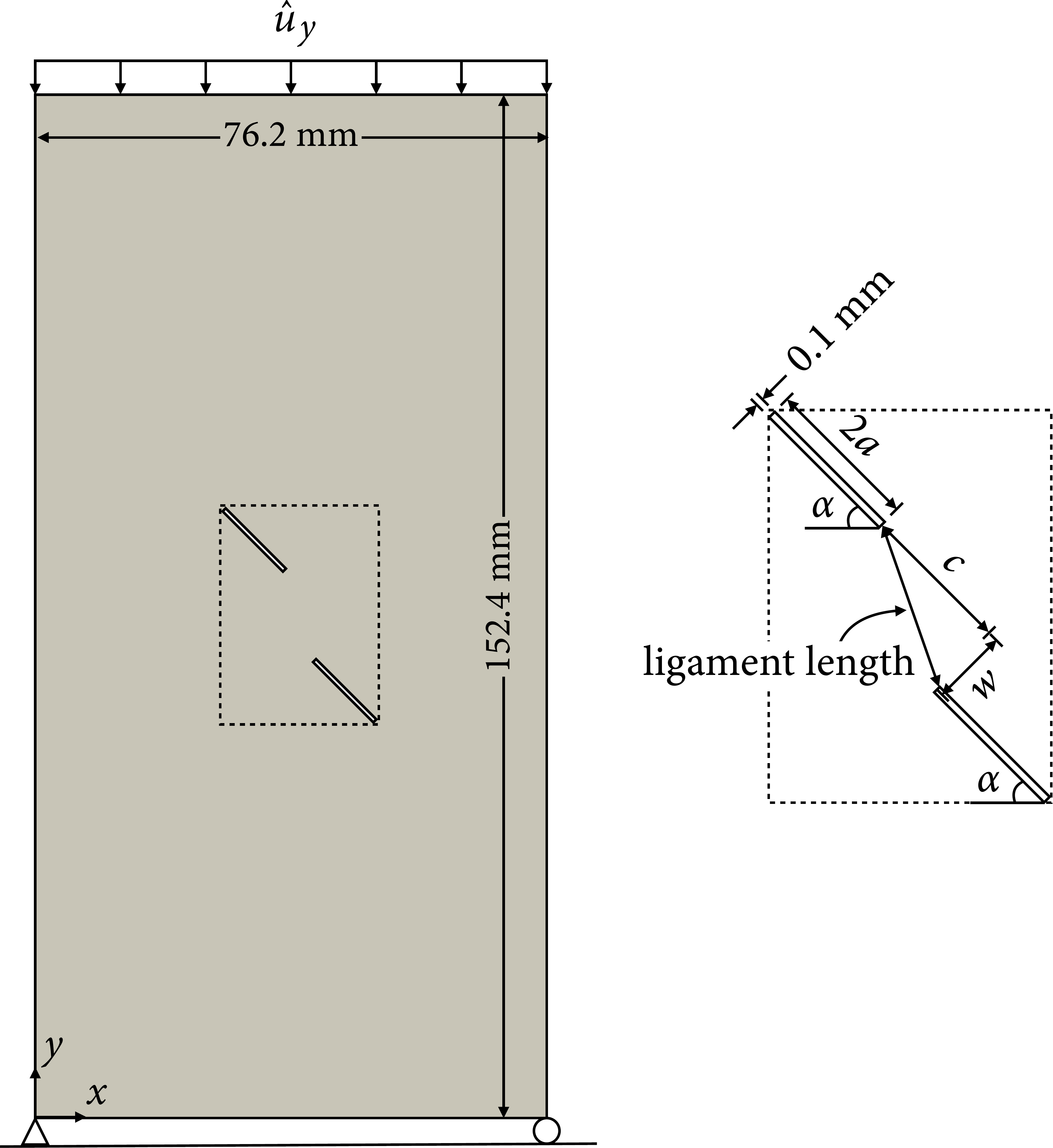}
    \caption{Cracking from double flaws: problem geometry and boundary conditions. The ligament length stands for the distance between the two flaws.}
    \label{fig:double-flaws-setup}
\end{figure}

Next, we simulate a variety of mixed-mode fracture processes in double-flawed specimens experimentally studied in Bobet and Einstein~\cite{bobet1998fracture} and Wong and Einstein~\cite{wong2009crack-a}.
Figure~\ref{fig:double-flaws-setup} depicts the general setup of specimens prepared according to the original experiments.
In all the specimens, the two flaws have the same length and aperture, 12.7 mm and 0.1 mm, respectively, with the continuity ($c$) of 12.7 mm.
By contrast, their inclination angle ($\alpha$) and spacing ($w$) are varied by specimens to trigger different types of cracking patterns under compression.
In this work, we particularly consider two cases of the inclination angle, $\alpha=45^\circ$ and $\alpha=60^\circ$, which manifested mixed-mode cracking patterns in the experiments.
Within the case of $\alpha=45^\circ$, we consider three sub-cases of flaw spacings: $w=0$, $w=a$, and $w=2a$, where $a$ denotes the half flaw length, $6.35$ mm.
Within the case of $\alpha=60^\circ$, we consider two sub-cases: $w=0$ and $w=a$.
As a result, we simulate a total of five cases.

In what follows, we compare our simulation results with the qualitative and quantitative data from Bobet and Einstein~\cite{bobet1998fracture}.
For the cases of zero spacing ($w=0$), Wong and Einstein~\cite{wong2009crack-a} later clarified the natures of cracks developed in gypsum specimens with the same flaw spacing.
For these cases, we will complement the qualitative experimental data by those provided in Wong and Einstein~\cite{wong2009crack-a}.

\begin{figure}[h!]
    \centering
    \includegraphics[width=1.0\textwidth]{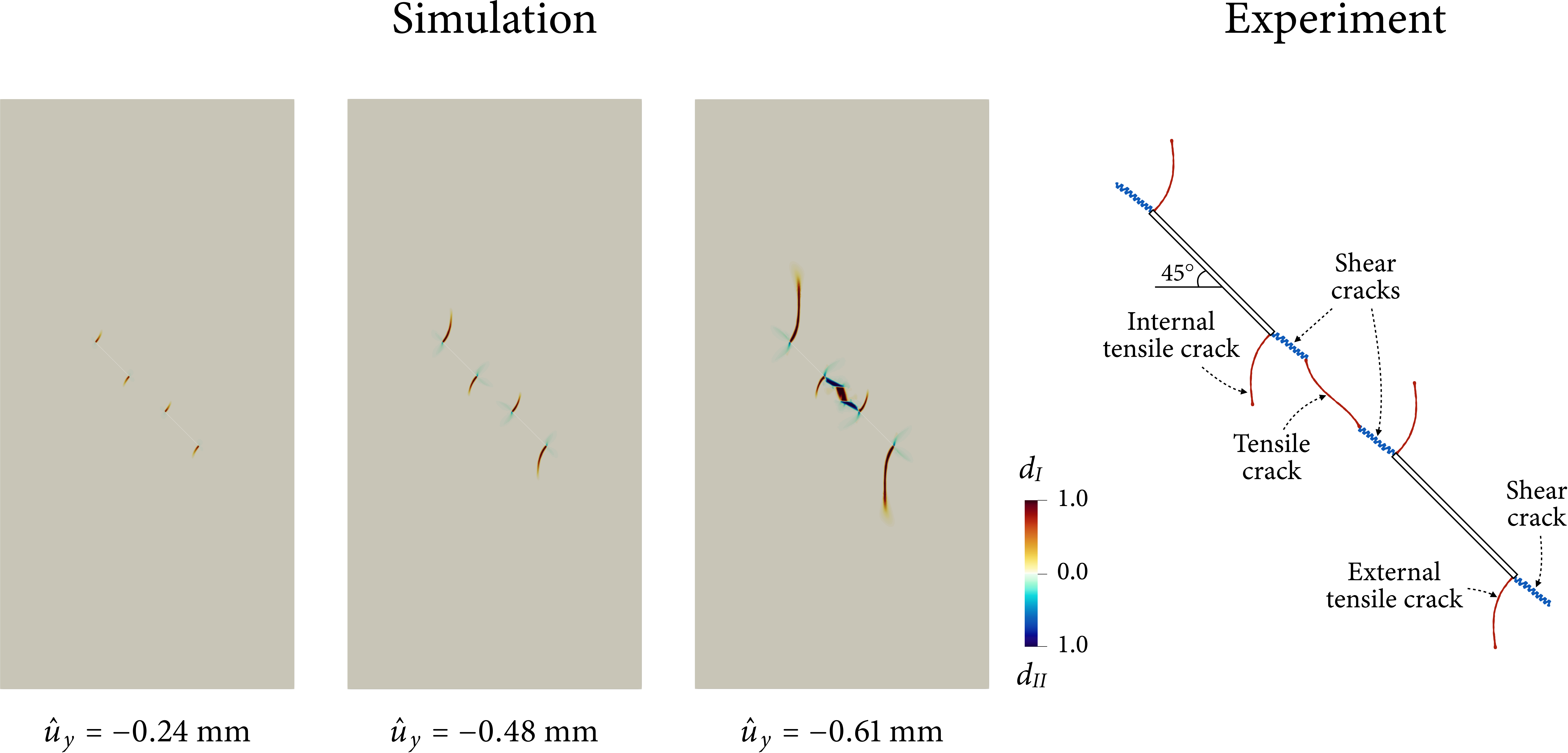}
    \caption{Cracking from double flaws with $\alpha = 45^\circ$ and $w = 0$ mm:
    simulation and experimental results. The experimental result is redrawn from Bobet and Einstein~\cite{bobet1998fracture} and Wong and Einstein~\cite{wong2009crack-a}.}
    \label{fig:45-0-2a-results}
\end{figure}

Figure~\ref{fig:45-0-2a-results} presents the simulation and experimental results when $\alpha=45^\circ$ and $w = 0$ mm.
Tensile wing cracks first develop from the flaw tips in a stable manner, and then shear damages grow in between the tips of the two flaws.
Eventually, the two flaws are coalesced by a mixed-mode crack, which consists of two coplanar shear cracks bridged by a tensile crack.
One can find that the simulation and experimental results are remarkably consistent in terms of the locations, shapes, and modes of the cracks.

\begin{figure}[h!]
    \centering
    \includegraphics[width=1.0\textwidth]{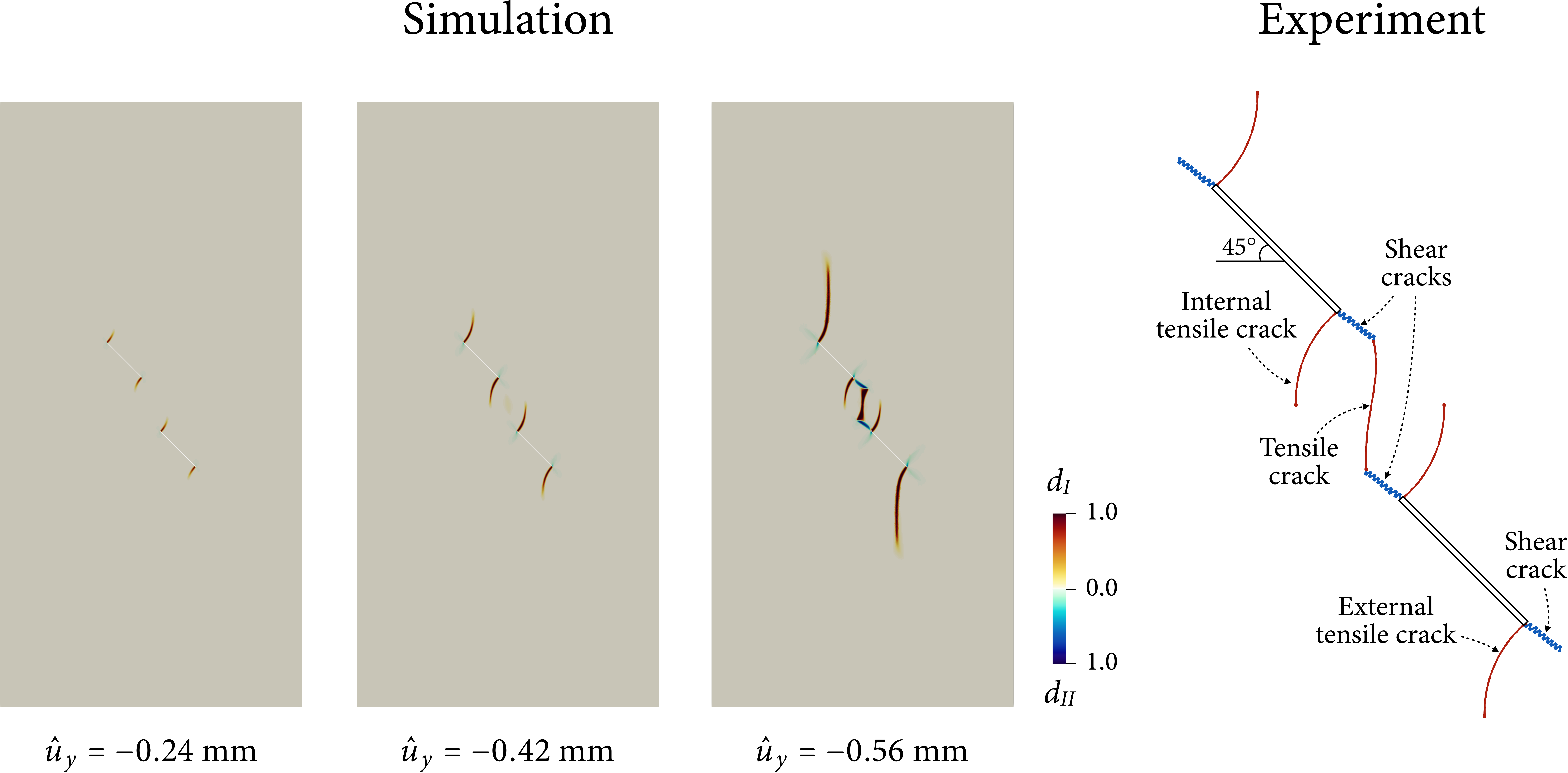}
    \caption{Cracking from double flaws with $\alpha = 45^\circ$ and $w = a = 6.35$ mm: simulation and experimental results. The experimental result is redrawn from Bobet and Einstein~\cite{bobet1998fracture}.}
    \label{fig:45-a-2a-results}
\end{figure}

Next, Fig.~\ref{fig:45-a-2a-results} shows and compares results from the simulation and the experiment when the spacing of the two flaws is increased to the half crack width, $a = 6.35$ mm.
The overall cracking process is similar to that in the previous case: tensile wing cracks followed by shear cracks and a secondary tensile crack which coalesce the two preexisting flaws.
Unlike the previous case, however, the coalescence crack in this case exhibits a zig-zag pattern.
This difference is also fully consistent with the experimental observations.

\begin{figure}[h!]
    \centering
    \includegraphics[width=1.0\textwidth]{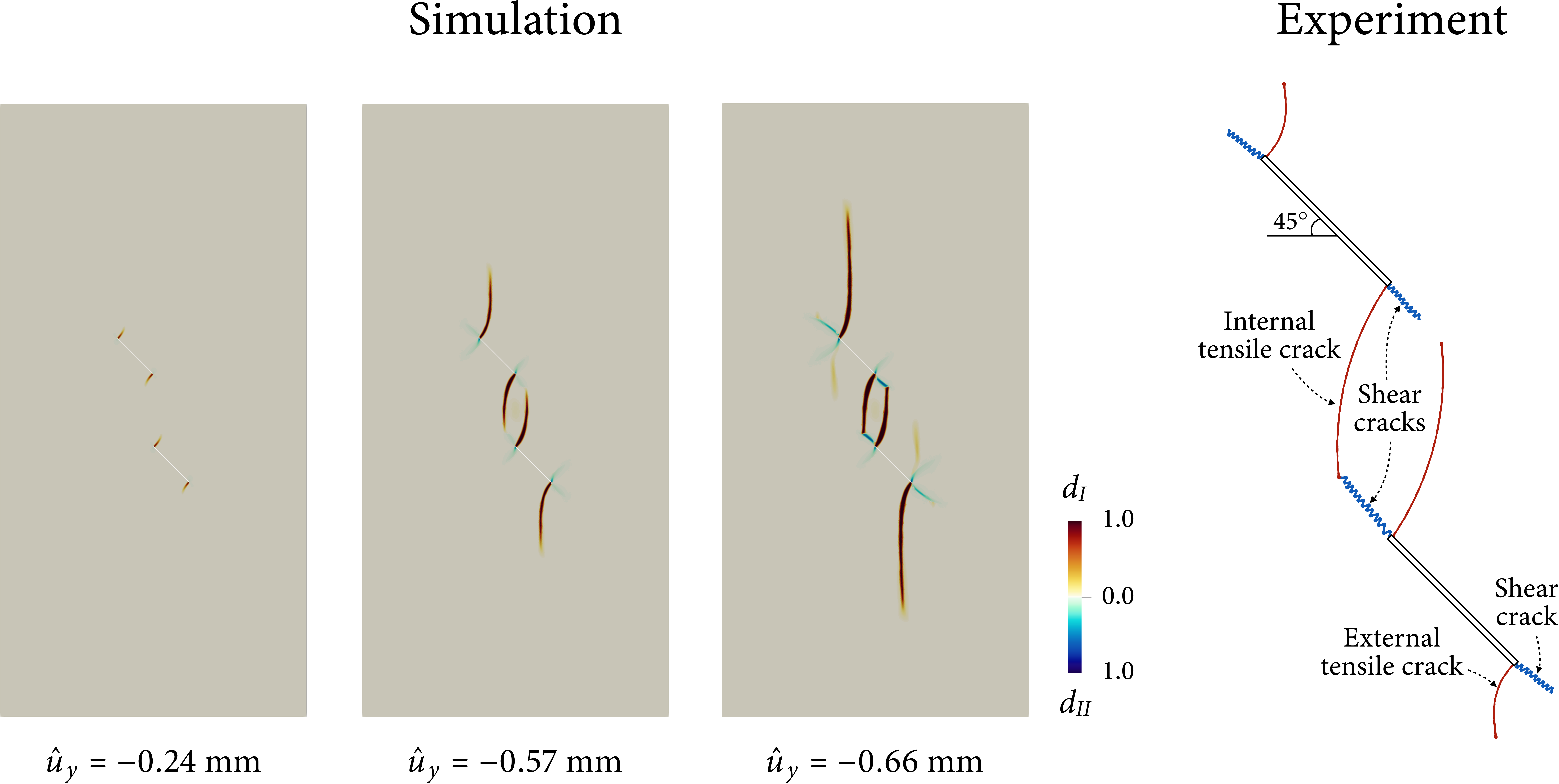}
    \caption{Cracking from double flaws with $\alpha = 45^\circ$ and $w = 2a = 12.7$ mm: simulation and experimental results. The experimental result is redrawn from Bobet and Einstein~\cite{bobet1998fracture}.}
    \label{fig:45-2a-2a-results}
\end{figure}

In Fig. \ref{fig:45-2a-2a-results}, we show the simulation and experimental results when the spacing is further increased to the crack width, $2a = 12.7$ mm.
The growth sequence of tensile wing cracks and shear cracks is the same as those in the previous two cases.
In the current case, however, the flaws are finally coalesced when a shear crack generated from one flaw links an internal wing crack from the other flaw.
This type of crack coalescence, which was not observed when $w/c < 1$, is also highlighted in the experimental study of Bobet and Einstein~\cite{bobet1998fracture}.
As can be seen, the proposed phase-field model can well capture this pattern transition as observed from the experiments.

\begin{figure}[h!]
    \centering
    \includegraphics[width=1.0\textwidth]{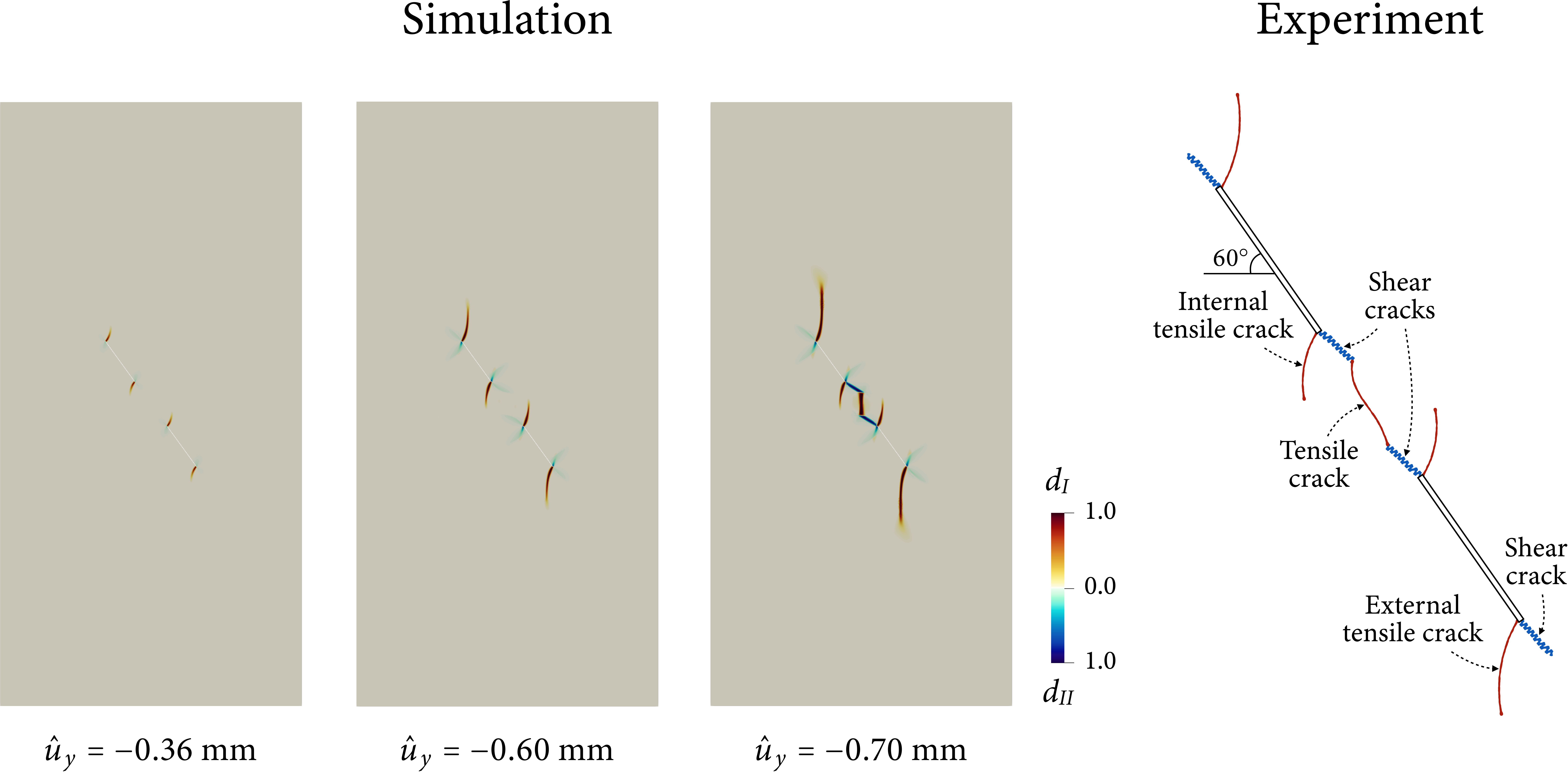}
    \caption{Cracking from double flaws with $\alpha = 60^\circ$ and $w = 0$ mm:
    simulation and experimental results. The experimental result is redrawn from Bobet and Einstein~\cite{bobet1998fracture} and Wong and Einstein~\cite{wong2009crack-a}.}
    \label{fig:60-0-2a-results}
\end{figure}

\begin{figure}[h!]
    \centering
    \includegraphics[width=1.0\textwidth]{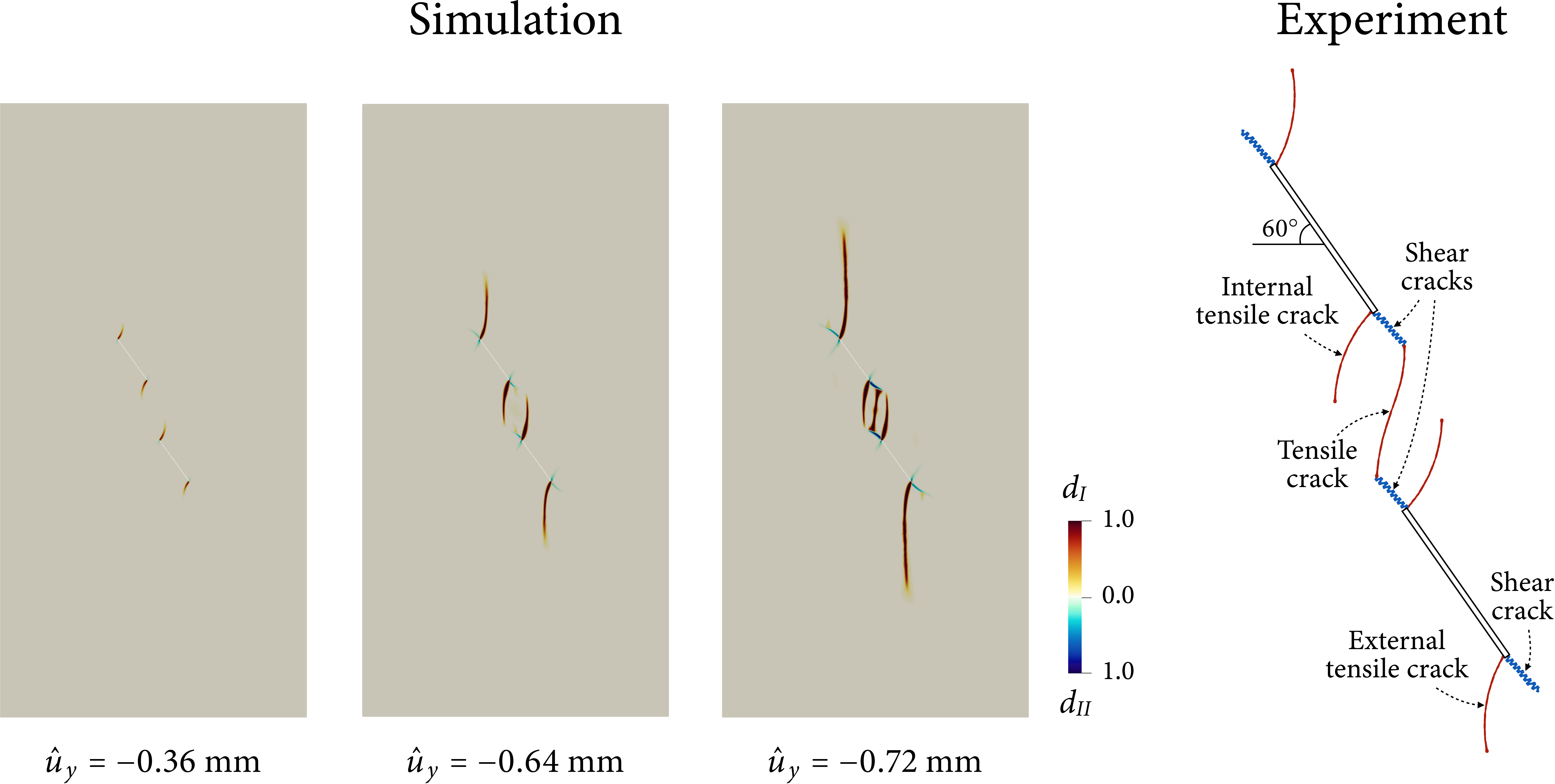}
    \caption{Cracking from double flaws with $\alpha = 60^\circ$ and $w = a = 6.35$ mm:
    simulation and experimental results. The experimental result is redrawn from Bobet and Einstein~\cite{bobet1998fracture}.}
    \label{fig:60-a-2a-results}
\end{figure}

Figures~\ref{fig:60-0-2a-results} and \ref{fig:60-a-2a-results} present how the simulation and experimental results become different as the flaw inclination angle is increased to $60^{\circ}$, when $w = 0$ mm and $w = a = 6.35$ mm, respectively.
The geometrical features of the secondary tensile cracks are changed, while the overall cracking patterns and sequences remain analogous to those of the cases of $\alpha = 45^\circ$.
The simulated coalescence crack in the case of $w = 0$ mm (Fig.~\ref{fig:60-0-2a-results}) is consistent with the experimental finding of Wong and Einstein~\cite{wong2009crack-a}, in that it is a mixed-mode crack consisting of two shear cracks developed from the inner flaw tips and a tensile crack bridging the shear cracks.
Also in the case of $w=a=6.35$ mm (Fig.~\ref{fig:60-a-2a-results}), a zig-zag coalescence pattern has emerged in both our simulation and the experiment of Bobet and Einstein~\cite{bobet1998fracture}.
Therefore, the simulation results are fully consistent with the experimental observations---from the geometry of cracks to the natures of tensile/shear cracks---under all of the flaw configurations.

\begin{figure}[h!]
    \centering
    \includegraphics[width=0.55\textwidth]{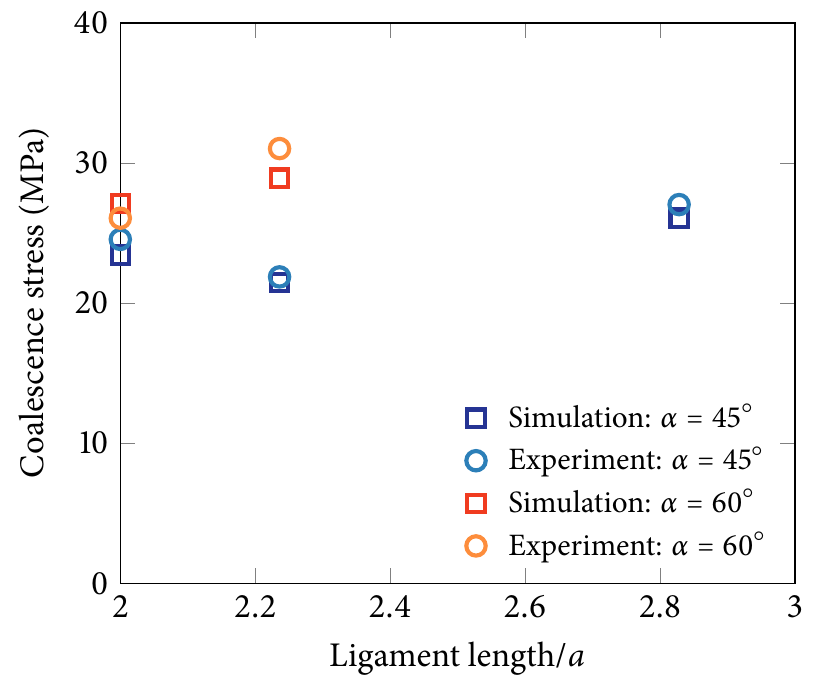}
    \caption{Cracking from double flaws: comparison of coalescence stresses in numerical simulation with the experimental data of Bobet and Einstein~\cite{bobet1998fracture}. (See Fig. \ref{fig:double-flaws-setup} for the definition of the ligament length.)}
    \label{fig:double-flaws-coalescence-stresses}
\end{figure}

Further, for quantitative validation, Fig. \ref{fig:double-flaws-coalescence-stresses} compares the coalescence stresses in the simulations and those measured in the experiments of Bobet and Einstein~\cite{bobet1998fracture}.
In all cases, the simulation results show excellent agreement with the experimental data.
Remarkably, the simulation results can well capture the increasing/decreasing trends of the coalescence stresses as observed from the experiments.

The numerical results in this section have demonstrated that the proposed phase-field model can not only reproduce mixed-mode fracture in the individual cases but also capture the transition of cracking patterns according to change in the flaw configurations.
The proposed model has thus been fully validated.
\smallskip

\revised{
\begin{remark}
  Some of the above experimental results have also been reproduced by the phase-field model for brittle mixed-mode fracture~\cite{zhang2017modification}.
  In the brittle model, however, the phase-field length parameter should be restricted to a specific value to match a prescribed tensile strength.
  Also, the shear strength of the brittle model cannot be controlled.
  Conversely, in the present quasi-brittle model, one can freely choose the length parameter because it does not affect the tensile and shear strengths of the material.
  Apart from its physical implications, this feature provides more flexibility to numerical modelers because the length parameter governs the discretization level in phase-field modeling.
\end{remark}}
\smallskip

\begin{remark}
  Besides validation, the above numerical results have demonstrated that the double-phase-field model allows us to naturally distinguish between tensile and shear cracks.
  This feature is invaluable to develop a better understanding of mixed-mode cracking processes in rocks.
  One main reason is that accurate experimental characterization of rock cracking processes requires a sophisticated technique (\eg~high speed imaging~\cite{wong2009using}) which is difficult, or even impossible, to be applied to rocks under in-situ stress conditions.
  The capability of providing physical insight into mixed-mode fracture without a sophisticated technique is a unique advantage of the double-phase-field formulation.
\end{remark}

\section{Closure}
\label{sec:closure}

We have developed a double-phase-field formulation for mixed-mode fracture in \revised{quasi-brittle} rocks,
employing two different phase fields to describe \revised{cohesive tensile fracture and frictional shear fracture} individually.
The formulation rigorously combines the two phase fields through three approaches:
(i) crack-direction-based decomposition of the strain energy into the tensile, shear, and pure compression parts,
(ii) contact-dependent calculation of the potential energy,
and (iii) energy-based determination of the dominant fracturing mode in each contact condition.
In this way, we have successfully coupled two types of phase-field models---one for cohesive tensile fracture and the other for frictional shear fracture---to model mixed-mode fracture in quasi-brittle rocks.
The double-phase-field model has been validated to reproduce a variety of mixed-mode fracturing processes in rocks, in both qualitative and quantitative senses.

\revised{Compared with the existing phase-field models for mixed-mode fracture in rocks,
the double phase-field model has two standout features.
First, it explicitly takes tensile and shear strengths as material parameters independent of the phase-field length parameter, unlike the existing models where the phase-field length controls the strengths.
This feature allows one to use experimentally-measured strengths directly without any restriction on the length parameter.}
Second, the double-phase-field model can simulate---and naturally distinguish between---tensile and shear fractures without complex algorithms.
This feature offers an exceptional opportunity to better understand rock cracking processes that are challenging, or even impossible, to be characterized by experiments alone.
Examples include crack growth and coalescence in rocks under true-triaxial stress conditions, which are much more difficult to be investigated experimentally than those under uniaxial/biaxial stress conditions.
We thus believe that the proposed model is an attractive option for both understanding and predicting mixed-mode fracture in rocks.

\section*{Acknowledgments}
The authors are grateful to Dr. Louis N.Y. Wong for sharing his experimental data and for helpful discussions regarding rock fracture.
The authors also thank Dr. Eric C. Bryant for his help with meshing.
This work was supported by the Research Grants Council of Hong Kong through Projects 17201419 and 27205918.
The first author also acknowledges financial support from a Hong Kong PhD Fellowship.

\bibliography{references}

\end{document}